\documentclass[aps,
    prb,
    amsmath,
    amssymb,
    floatfix,
    reprint
    ]{revtex4-2}
\usepackage{graphicx}
\usepackage{physics}
\usepackage{booktabs}
\usepackage{amsthm,amssymb,mathrsfs}
\usepackage{tensor}
\usepackage{bm}

\usepackage{hyperref} % For hyperlinks in the PDF
\hypersetup{colorlinks, allcolors=blue}

\usepackage{wasysym}
\usepackage{xargs} 
\usepackage{xcolor}

\DeclareMathAlphabet{\mathdutchcal}{U}{dutchcal}{m}{n}
\SetMathAlphabet{\mathdutchcal}{bold}{U}{dutchcal}{b}{n}
\DeclareMathAlphabet{\mathdutchbcal}{U}{dutchcal}{b}{n}

\usepackage[inline]{enumitem}

\usepackage{enumitem}
\setlist[itemize]{leftmargin=*}
\setlist[enumerate]{leftmargin=*}
\let\temp\phi
\let\phi\varphi
\let\varphi\temp

\newcommand{\s}{\mathcal{S}}

\usepackage{scalerel,stackengine}

\newcommand{\iu}{\mathrm{i}} % imaginary unit
\newcommand{\e}{\mathrm{e}} % exponent

\newcommand{\DD}{\mathdutchcal{D}} % Path integral measure
\newcommand{\D}{\mathrm{d}}
\newcommand{\hc}{\mathrm{h.c.}}
\newcommand{\sign}{\operatorname{sgn}} % sign function

\newcommand{\snn}[2]{\sum_{\langle #1, #2 \rangle}}

\providecommand{\CC}{\mathbb{C}}

\providecommand{\ZZ}{\mathbb{Z}}

\let\v\relax % 
\newcommand{\v}[1]{\ensuremath{\mathbf{#1}}} % vektor
\makeatletter
\newcommand*{\coloneqq}{\mathrel{\rlap{%
			\raisebox{0.28ex}{$\m@th\cdot$}}%
		\raisebox{-0.28ex}{$\m@th\cdot$}}%
	=}
\newcommand*{\eqqcolon}{=\mathrel{\rlap{%
			\raisebox{0.28ex}{$\m@th\cdot$}}%
		\raisebox{-0.28ex}{$\m@th\cdot$}}%
	}
\makeatother

\makeatletter
\newcommand*{\rom}[1]{\expandafter\@slowromancap\romannumeral #1@}
\makeatother

\DeclareSymbolFont{cmbrightop}{OT1}{cmbr}{m}{n}
\DeclareMathSymbol{\sfPsi}{\mathalpha}{cmbrightop}{9}

\begin{document}
	
	\title{Topological superconductivity induced by a Kitaev spin liquid}
	
	\author{Sondre Duna Lundemo} 
	\affiliation{Center for Quantum Spintronics, Department of Physics, Norwegian University of Science and Technology, NO-7491 Trondheim, Norway}
 
	\author{Asle Sudb\o}
	\email[Corresponding author: ]{asle.sudbo@ntnu.no}
	\affiliation{Center for Quantum Spintronics, Department of Physics, Norwegian University of Science and Technology, NO-7491 Trondheim, Norway}

\date{\today} 

\begin{abstract}
	We study the effective low-energy fermionic theory of the Kondo-Kitaev model to leading order in the Kondo coupling.
	Our main goal is to understand the nature of the superconducting instability induced in the proximate metal due to its coupling to spin fluctuations of the spin liquid.
    The special combination of the low-energy modes of a graphene-like metal and the form of the interaction induced by the Majorana excitations of the spin liquid furnish chiral superconducting order with $p_x + \iu p_y$ symmetry. 
    Computing its response to a $\mathrm{U}(1)$ gauge field moreover shows that this superconducting state is topologically non-trivial, characterized by a first Chern number of $\pm 2$.
\end{abstract}

\maketitle 

%\tableofcontents

\section{Introduction}

Understanding, and suggesting platforms for topological superconductivity (TSC) has become a central problem in condensed matter physics, largely motivated by its possible application in topological quantum computing \cite{nayakNonAbelianAnyonsTopological2008,menardIsolatedPairsMajorana2019,satoTopologicalSuperconductorsReview2017}.
Since materials that support this phase intrinsically are rare in nature, the search for TSC has mainly been restricted to interfaces between exotic magnets and conventional superconductors \cite{zlotnikovAspectsTopologicalSuperconductivity2021,nakosaiTwodimensionalWaveSuperconducting2013,chenMajoranaEdgeStates2015,rexMajoranaBoundStates2019}.
In particular, a combination of strong spin-orbit coupling and Zeeman fields is conjectured to induce TSC in the superconductors of these proposed systems \cite{menardIsolatedPairsMajorana2019}.
More recently, a system comprised of a skyrmion crystal interfaced with a normal metal was shown theoretically to produce TSC at the interface, effectively removing the indispensable component of previous suggestions, namely conventional superconductors \cite{maelandTopologicalSuperconductivityMediated2023}.
%\textcolor{red}{
The model of the present work is similar in spirit, in the sense that neither Zeeman fields nor conventional superconductors are required for the spin fluctuations to induce TSC.
%}

The study of quantum spin liquid (QSL) states of spin systems \cite{andersonResonatingValenceBonds1973,kalmeyerEquivalenceResonatingvalencebondFractional1987,wenChiralSpinStates1989}, and particularly the construction of exactly solvable Hamiltonians featuring QSL ground states \cite{kitaevAnyonsExactlySolved2006}, has inspired the search for TSC.
QSL states are exotic ground states of spin systems that do not feature long-range magnetic order, but rather display topological order and host fractionalized excitations \cite{wenMeanfieldTheorySpinliquid1991a,wenEffectiveTheoryPbreaking1989}.
While most of these properties are poorly understood within traditional perturbative approaches, there are fortunate rare cases where we are guided by exact solutions. 
One example of this is the Kitaev honeycomb model \cite{kitaevAnyonsExactlySolved2006}, which consists of localized spins on a honeycomb lattice interacting through link-dependent Ising interactions.
For such systems coupled to itinerant fermions, it is natural to ask whether the associated spin fluctuations can induce superconductivity in the metal, and if so, to what extent this state inherits the topological nature of the parent QSL.
Following recent developments in the theory of Kitaev materials that couple spin models with (Kitaev) QSL ground states to conduction electrons \cite{seifertFractionalizedFermiLiquids2018b,choiTopologicalSuperconductivityKondoKitaev2018,decarvalhoOddfrequencyPairDensity2021,colemanSolvable3DKondo2022,tsvelikOrderFractionalizationKitaevKondo2022}, the present work aims to answer these questions.

\begin{figure}[htb]
	\centering
	\includegraphics[width=0.9\columnwidth]{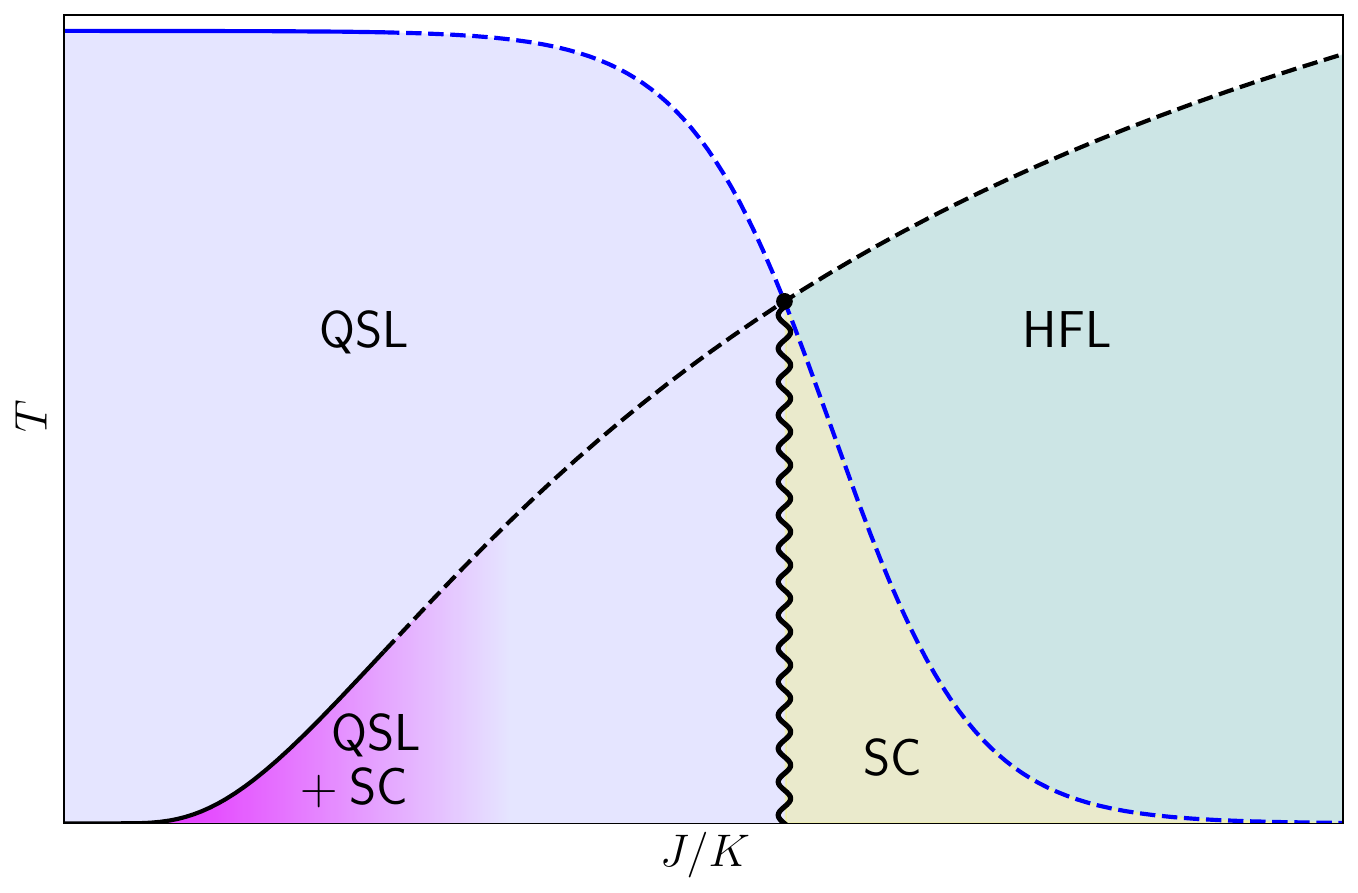}
	\caption{Schematic phase diagram of the system, extended from Refs.~\cite{senthilFractionalizedFermiLiquids2003,seifertFractionalizedFermiLiquids2018b}. The quantum spin liquid phase is denoted by $\mathsf{QSL}$ and the superconducting state where superconductivity of conduction electrons coexists with the spin-liquid phase is denoted by $\mathsf{QSL}+\mathsf{SC}$. 
	Beyond the perturbative regime of $J/K$ one finds a heavy Fermi liquid ($\mathsf{HFL}$) and a superconducting phase ($\mathsf{SC}$), both due to the Kondo effect as described by Ref.~\cite{senthilFractionalizedFermiLiquids2003}. 
	The wiggly line represents a first-order transition separating the fractionalized Fermi-liquid phase from the non-fractionalized one, due to competing order parameters.}
	\label{fig:pd}
\end{figure}
To this end, we consider a system comprised of localized spins on a honeycomb lattice governed by the Kitaev interaction with interaction strength $K$, itinerant electrons on a proximate honeycomb lattice, and couple these through a Kondo interaction with interaction strength $J$ (see Eq.~\eqref{eq:total_H}). 
For $J=0$ and for sufficiently low temperatures, the system exhibits the QSL phase. 
The perturbative regime of $J/K$ finite but small is continuously connected to the $J=0$ limit \cite{senthilFractionalizedFermiLiquids2003,seifertFractionalizedFermiLiquids2018b}.
However, it is conceivable that a finite, small $J$ will induce an attractive interaction between the conduction electrons, facilitating a superconducting instability of the Fermi sea. 
This is analogous to the mechanism by which magnons of a ferro- or antiferromagnet Kondo-coupled to a conductor mediates superconductivity \cite{kargarianAmpereanPairingSurface2016,Rohling_2018,hugdalMagnoninducedSuperconductivityTopological2018,erlandsenEnhancementSuperconductivityMediated2019a,erlandsenMagnonmediatedSuperconductivitySurface2020,thingstadEliashbergStudySuperconductivity2021}, except that the mediator, in the present case, is the fractionalized excitations of the spin-liquid.
Increasing $J/K$ beyond the perturbative regime $J/K\ll1$, conduction electrons will hybridize with the localized spins and form Kondo singlets \cite{colemanKondostabilisedSpinLiquids1989,chatterjeeSuperconductivityConfinementTransition2016}. 
At sufficiently low temperatures, the metal will turn superconducting whereas at higher temperatures it will be a heavy Fermi liquid.
The transition between the QSL phase and this superconducting phase will generically be separated by a first-order transition, as it originates with the competition between two orders \cite{imryStatisticalMechanicsCoupled1975a,bruceCoupledOrderParameters1975a,calabreseMulticriticalPhenomenaMathrm2003a}.
The phase diagram of this system is schematically illustrated in Fig.~\ref{fig:pd}.
The previous works concerned with the superconductivity of this model chiefly focus on the phase denoted by $\mathsf{SC}$ in this figure \cite{seifertFractionalizedFermiLiquids2018b,choiTopologicalSuperconductivityKondoKitaev2018,decarvalhoOddfrequencyPairDensity2021}.
The regime we focus on is illustrated as the pink region, fading over into a regime inaccessible to our study which is schematically extended by dashed lines to qualitatively agree with those of \cite{senthilFractionalizedFermiLiquids2003,seifertFractionalizedFermiLiquids2018b}.

\section{The Kondo-Kitaev Model}

We consider a honeycomb lattice $\Lambda \ni i$ with lattice constant $a$. 
%The honeycomb lattice is bipartite, being the direct sum of two triangular lattices.
To each vertex of this bipartite lattice, we associate a fermionic degree of freedom with creation and annihilation operators $c^{\dagger}_{i\sigma}$ and $c_{i\sigma}^{\mathstrut}$ obeying the canonical anticommutation relations
\begin{equation}\label{eq:cars}
	\begin{split}
		&\{c_{i\alpha}^{\mathstrut}, c_{j\beta}^{\dagger}\} = \delta_{ij} \delta_{\alpha\beta} \quad \text{and} \\
		&\{c_{i\alpha}^{\dagger},c_{j\beta}^{\dagger}\} = 0 = \{c_{i\alpha}^{\mathstrut},c_{j\beta}^{\mathstrut}\}
	\end{split}
\end{equation}
and a spin-$1/2$ degree of freedom, whose components satisfy
\begin{equation}\label{eq:spin_algebra}
	[S^{\mathsf{a}}_{i},S^{\mathsf{b}}_{j}] = \iu \delta_{ij} \epsilon^{\mathsf{a} \mathsf{b} \mathsf{c}} S_{i}^{\mathsf{c}}, \quad \,\mathsf{a},\mathsf{b},\mathsf{c} \in \{ x,y,z\},\end{equation}
with summation over repeated indices.
In the following, we use Latin letters $ijk\dots$ for lattice points, Greek letters $\alpha\beta\gamma\dots$ for spin indices of itinerant fermions, and sans serif letters $\mathsf{a}\mathsf{b}\mathsf{c}\dots$ for components of the localized spin operators and link indices (to be introduced shortly).

\begin{figure}[htb]
	\centering
	\includegraphics[width=0.8\columnwidth]{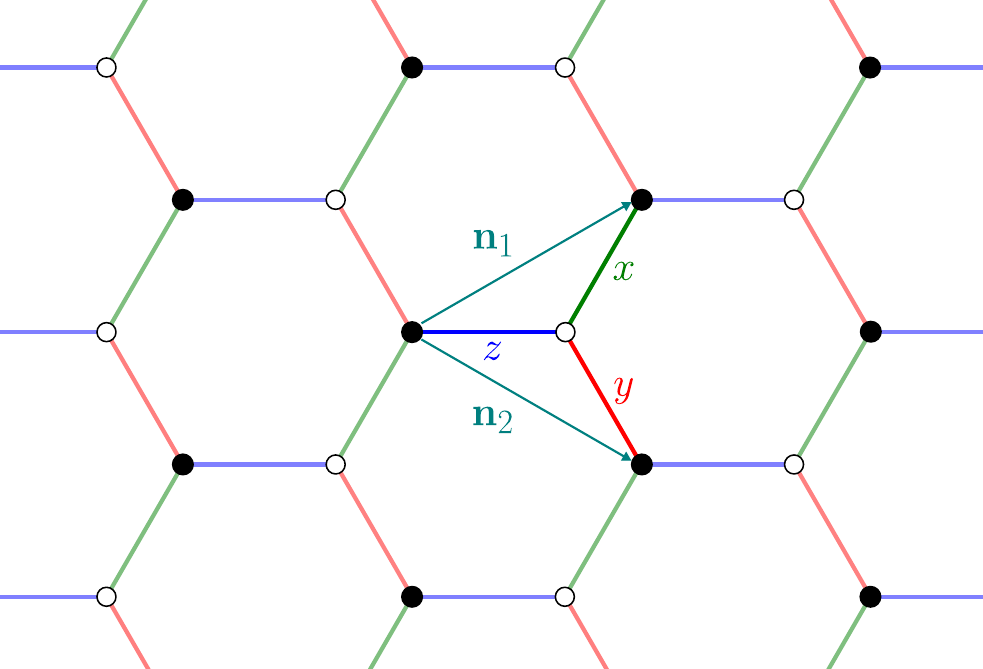}
	\caption{The honeycomb lattice with the $x$-,$y$- and $z$-links colored in green, red, and blue respectively. The filled (hollow) lattice sites belong to the A (B) sublattice, and the vectors $\v{n}_{1,2} \coloneqq a (\sqrt{3},\pm 1)^{\mathsf{T}}/2$ are the lattice translation vectors of the hexagonal lattice.}	
	\label{fig:lattice}
\end{figure}

The Hamiltonian of the Kondo--Kitaev model is given by
\begin{subequations}\label{eq:total_H}
	\begin{align}
		H &\coloneqq H_{\mathrm{el}} + H_K + H_J, \\
		\intertext{where}
		H_{\mathrm{el}} &\coloneqq -t \snn{i}{j} \sum_{\sigma} c^{\dagger}_{i\sigma} c^{\mathstrut}_{j\sigma} - \mu\sum_{i\in\Lambda} \sum_{\sigma} c^{\dagger}_{i\sigma} c^{\mathstrut}_{i\sigma} \label{eq:H_el} \\
		H_K &\coloneqq - K \sum_{\mathsf{a}=1}^{3}\sum_{\langle i, j \rangle_{\mathsf{a}}} S^{\mathsf{a}}_i S^{\mathsf{a}}_{j} \label{eq:H_kitaev}\\
		H_J &\coloneqq + \frac{J}{2} \sum_{i\in\Lambda} \sum_{\alpha\beta}\sum_{\mathsf{a}=1}^{3} c^{\dagger}_{i\alpha} \sigma^{\mathsf{a}}_{\alpha\beta} c^{\mathstrut}_{i\beta} S^{\mathsf{a}}_{i}. \label{eq:H_kondo}
	\end{align}
\end{subequations}
The symbol $\langle i,j \rangle_{\mathsf{a}}$ denotes the lattice point pair $i$ and $j$ corresponding to the $\mathsf{a}$ link of the honeycomb lattice, as illustrated in Fig.~\ref{fig:lattice}. 
The Kitaev interaction assigns an Ising interaction on link $\mathsf{a}$ along the direction $\mathsf{a}$ in spin space.

Before studying the complete Kondo-Kitaev model we will consider the mean-field ground state of $H_{\mathrm{K}}$ on its own.
To this end, we employ a Majorana representation of the localized spins,  discussed extensively in the literature \cite{youDopingSpinorbitMott2012,seifertFractionalizedFermiLiquids2018b}.
We briefly revisit some properties of this representation for completeness and refer back to these references for details.

\subsection{Majorana representation of localized spins}

For studying spin liquids, we start with a \textit{slave fermion} representation of the spin operators in terms of fermionic creation and annihilation operators $f_{i\sigma}^{\dagger}$ and $f_{i\sigma}^{\mathstrut}$ as \cite{Abrikosov1965,affleckSUGaugeSymmetry1988}
\begin{equation}\label{eq:slave_fermions}
	\v{S}_{i} = \frac{1}{2} f^{\dagger}_{i\alpha} \bm{\sigma}_{\alpha\beta}^{\mathstrut} f_{i\beta}^{\mathstrut},
\end{equation}
which are constrained to satisfy $n_{i\uparrow} + n_{i\downarrow} = 1$ at the operator level. By arranging these operators in a matrix 
\begin{equation}\label{eq:Fspin}
	F_{i} \coloneqq \begin{pmatrix}
		f_{i\uparrow}^{\mathstrut} & - f_{i\downarrow}^{\dagger} \\
		f_{i\downarrow}^{\mathstrut} & f_{i\uparrow}^{\dagger}, 
	\end{pmatrix}, 
\end{equation}
one can translate the representation into one of \textit{Majorana fermions}, satisfying $\left(\chi^{\mu}_i\right)^{\dagger} = \chi^{\mu}_{i}$ and the anti-commutation relations 
\begin{equation}
	\left\{ \chi^{\mu}_{i}, \chi_{j}^{\nu} \right\} = \delta^{\mu\nu} \delta_{ij},
\end{equation}
where $\mu,\nu=0,\dots,3$ \cite{kitaevAnyonsExactlySolved2006,tsvelikNewFermionicDescription1992}. 
The correspondence is established by letting \cite{youDopingSpinorbitMott2012}
\begin{equation}
	F_ {i} = \frac{1}{\sqrt{2}} \left( \sigma^0 \chi^{0}_ {i} + \sum_{\mathsf{a} = 1}^{3} \iu \sigma^{\mathsf{a}} \chi^{\mathsf{a}}_{i} \right).
\end{equation}
Combining this expression with Eqs.~\eqref{eq:slave_fermions} and \eqref{eq:Fspin}, one finds that 
\begin{equation}\label{eq:spin_majorana}
	S^{\mathsf{a}}_i = \frac{1}{4} \tr F_{i}^{\dagger} \sigma^{\mathsf{a}} F_{i} = \frac{\iu}{2} \left( \chi^{0}_{i}\chi_{i}^{\mathsf{a}} - \frac{1}{2} \epsilon^{\mathsf{a}\mathsf{b}\mathsf{c}} \chi_{i}^{\mathsf{b}} \chi_{i}^{\mathsf{c}} \right),
\end{equation}
while the single-occupancy constraint can be cast in the form 
\begin{equation}\label{eq:isospin_majorana}
	J^{\mathsf{a}}_{i} \coloneqq -\frac{\iu}{2} \left( \chi^{0}_{i}\chi^{\mathsf{a}}_{i} + \frac{1}{2}\epsilon^{\mathsf{abc}}\chi^{\mathsf{b}}_{i}\chi^{\mathsf{c}}_{i} \right) = 0.
\end{equation}
In the above equation, we introduced the \textit{isospin} $J^{\mathsf{a}}$, and the constraint identifies the physical Hilbert space with that of \textit{isospin singlets}. Following Ref.~\cite{seifertFractionalizedFermiLiquids2018b}, we write Eqs.~\eqref{eq:spin_majorana} and \eqref{eq:isospin_majorana} in matrix form 
\begin{equation}\label{eq:spin_and_isospin}
	\v{S}_{i} = \frac{\iu}{4}\chi^{\mu}_{i} \v{M}_{\mu\nu} \chi^{\nu}_{i} \quad \text{and}\quad \v{J}_{i} = \frac{\iu}{4}\chi^{\mu}_{i} \v{G}_{\mu\nu} \chi^{\nu}_{i},
\end{equation}
where the $\mathrm{SO}(4)$ matrices are given by
\begin{align*}
		M^{1} &\coloneqq \sigma^{3} \otimes \iu \sigma^2  &&M^2 \coloneqq \iu \sigma^2 \otimes \sigma^0 &&M^3 \coloneqq \sigma^1 \otimes \iu\sigma^2, \\
		G^{1} &\coloneqq -\sigma^{0} \otimes \iu \sigma^2  &&G^2 \coloneqq -\iu \sigma^2 \otimes \sigma^3  &&G^3 \coloneqq -\iu\sigma^2 \otimes \sigma^1.
\end{align*}
Inspired by Kitaev's exact solution \cite{kitaevAnyonsExactlySolved2006} and assuming isospin-singlet Majoranas, we can modify the spin operator to
\begin{equation}\label{eq:kitaev_rep}
	S^{\mathsf{a}}_{\mathrm{K},i} = \frac{\iu}{4} \chi^{\mu}_{i} \left[ M^{\mathsf{a}} -  G^{\mathsf{a}}\right]_{\mu\nu} \chi^{\nu}_{i} = \iu \chi^{0}_{i}\chi^{\mathsf{a}}_{i}.
\end{equation}

\subsection{Functional Integral Formulation}

In the functional-integral representation, we give the anticommuting operators imaginary time dependence and replace them with Grassmann-valued fields
\begin{equation*}
    c_{i\sigma}^{\mathstrut}(\tau) \to \psi_{i\sigma}(\tau) \quad \text{and} \quad c_{i\sigma}^{\dagger}(\tau) \to \bar{\psi}_{i\sigma}(\tau),
\end{equation*}
and likewise for the Majorana operators $\chi_{i}^{\mu}(\tau)$, except that we do not distinguish between the symbol used for the operator and the Grassmann field in this case. 

For the moment, we use the general Majorana spin representation given in Eq.~\eqref{eq:spin_and_isospin}, such that the Kitaev interaction  is given by
\begin{align}
    \s_{\mathrm{K}} &= - K \int_{0}^{\beta}\D\tau \sum_{\mathsf{a}=1}^{3}\sum_{\langle i,j \rangle_{\mathsf{a}}} \frac{1}{4^2} M^{\mathsf{a}}_{\mu\nu} M^{\mathsf{a}}_{\rho\sigma} \iu \chi^{\mu}_{i} \chi^{\nu}_{i} \iu \chi^{\rho}_{j} \chi^{\sigma}_{j} \notag \\
    &\equiv - \int_{0}^{\beta}\D\tau \sum_{\mathsf{a}=1}^{3}\sum_{\langle i,j \rangle_{\mathsf{a}}} V^{\mathsf{a}}_{\mu\nu\rho\sigma} \iu \chi^{\mu}_{i} \chi^{\nu}_{j} \iu \chi^{\rho}_{j} \chi^{\sigma}_{i}, 
\end{align}
with
\begin{equation}
    V^{\mathsf{a}}_{\mu\nu\rho\sigma} \coloneqq \frac{K}{16} M^{\mathsf{a}}_{\mu\sigma} M^{\mathsf{a}}_{\nu\rho}.
\end{equation}
The quartic Majorana term is decoupled via a Hubbard-Stratonovich transformation by introducing a real auxiliary field $\Phi^{\mu\nu}_{ij}$ alongside a measure $\DD \Phi$ normalized so that
\begin{equation}\label{Resolve}
    1 = \int \DD \Phi \exp\left( - \int_{0}^{\beta}\D \tau \sum_{\mathsf{a}=1}^{3}\sum_{\langle i,j\rangle_{\mathsf{a}}} \Phi_{ij}^{\mu\nu} \left(V^{\mathsf{a}}\right)^{-1}_{\mu\nu\rho\sigma} \Phi_{ji}^{\rho\sigma} \right).
\end{equation}
For the moment, we keep the inverse $(V^{\mathsf{a}})^{-1}$ unspecified, but note that it satisfies 
\begin{equation*}
(V^{\mathsf{a}})^{-1}_{\mu\nu\rho\sigma}V^{\mathsf{a}}_{\rho\sigma\mu'\nu'} = \delta_{\mu\mu'} \delta_{\nu\nu'}.
\end{equation*}
Regarding the pair of indices $\mu\nu$ as a composite vector index allows us to employ a matrix notation for the action of the auxiliary field, namely
\begin{equation}\label{eq:s_phi}
     \s_{\Phi} \coloneqq \int_{0}^{\beta}\D \tau \sum_{\mathsf{a}=1}^{3}\sum_{\langle i,j\rangle_{\mathsf{a}}} \bm{\Phi}_{ij}^{\mathsf{T}} \left(V^{\mathsf{a}}\right)^{-1} \bm{\Phi}_{ji}.
\end{equation} 
Using Eq.~\eqref{Resolve} and performing a linear shift in the $\bm{\Phi}$ fields 
\begin{equation}
    \begin{split}
        	\Phi_{ij}^{\mu\nu} &\mapsto \Phi_{ij}^{\mu\nu} - V^{\mu\nu\rho\sigma} \iu\chi^{\rho}_{i}\chi^{\sigma}_{j} \\
         	(\Phi_{ij}^{\mathsf{T}})^{\mu\nu}  &\mapsto (\Phi_{ij}^{\mathsf{T}})^{\mu\nu} -  V^{\rho\sigma\mu\nu} \iu\chi^{\rho}_{i}\chi^{\sigma}_{j},
    \end{split}
\end{equation}
we eliminate the quartic interaction between the Majorana fermions in favor of linear couplings between Majorana bilinears and the auxiliary bosons.

To implement the isospin-singlet constraint $J^{\mathsf{a}} = 0$, we introduce a fluctuating bosonic field $\lambda$ through the Gutzwiller projection \cite{colemanKondostabilisedSpinLiquids1989}
\begin{align}
    \delta\left(J^{\mathsf{a}}\right) &= \int \frac{\DD \lambda}{2\pi} \exp\left(- \frac{\iu}{2}\sum_{i\in \Lambda}\int_{0}^{\beta} \D\tau \sum_{\mathsf{a}} \chi^{\mu}_{i} \lambda_i^{\mathsf{a}} G_{\mu\nu}^{\mathsf{a}} \chi^{\nu}_{i} \right) \notag \\
    &\equiv \int \DD W \exp\left( -\frac{\iu}{2}\sum_{i\in \Lambda}\int_{0}^{\beta} \D\tau \bm{\chi}_{i}^{\mathsf{T}} W_{i} \bm{\chi}_{i} \right),
\end{align}
where $W_i \coloneqq \lambda^{\mathsf{a}}_{i} G^{\mathsf{a}}$ is an $\mathrm{SU}(2)$-valued auxiliary field. 
The resulting Hubbard-Stratonovich transformed action of the system reads
\begin{subequations}
\begin{equation}
    \s\left[\Phi,\chi,W,\bar{\psi},\psi\right] = \s_{\Phi} + \s_{\chi} + \s_{\Phi\chi} + \s_{\psi} + \s_{\chi\psi},
\end{equation}
with $\s_{\Phi}$ given in Eq.~\eqref{eq:s_phi} and
\begin{align}
    \s_{\chi} &\coloneqq  \frac{1}{2} \int_{0}^{\beta} \D \tau  \sum_{i\in \Lambda} \chi^{\mu}_{i} \left[ \delta^{\mu\nu} \partial_{\tau} + \iu W^{\mu\nu}_{i} \right] \chi^{\nu}_{i} \\
    \s_{\Phi\chi} &\coloneqq -\int_{0}^{\beta} \D\tau \sum_{\mathsf{a}=1}^{3}\sum_{\langle i,j\rangle_{\mathsf{a}}} 2 \iu \chi^{\mu}_{i} \chi^{\nu}_{j} \Phi^{\mu\nu}_{ji} \\
    \s_{\psi} &\coloneqq \int_{0}^{\beta}\D\tau \sum_{i,j \in \Lambda} \bar{\psi}_{i} \left[ \delta_{ij}(\partial_{\tau} - \mu) - t \delta_{i+\bm{\delta},j} \right] \psi_{j} \\
    \s_{\chi\psi} &\coloneqq  \frac{J}{2}\int_{0}^{\beta}\D \tau \sum_{i \in \Lambda} \sum_{\mathsf{a}=1}^{3} \bar{\psi}_{i\alpha} \sigma^{\mathsf{a}}_{\alpha\beta} \psi_{i\beta} \iu \chi^{0}_{i} \chi^{\mathsf{a}}_{i},
\end{align}
\end{subequations}
where we have used Eq.~\eqref{eq:kitaev_rep} directly in $S_{\chi\psi}$, and denoted the nearest neighbors of $i\in\Lambda$ by $i+\bm{\delta}$.
A justification for the former will be provided in the saddle-point analysis.

\subsection{Saddle-point analysis for $J=0$}

For completeness and to establish connections to previous works \cite{kitaevAnyonsExactlySolved2006,youDopingSpinorbitMott2012,seifertFractionalizedFermiLiquids2018b}, we set $J=0$ for the moment and solve the saddle-point equations of $\s_{\chi} + \s_{\Phi\chi}$.
In this calculation, we leave the spin representation on the form given in Eq.~\eqref{eq:spin_and_isospin} and connect the results to the representation Eq. \eqref{eq:kitaev_rep} towards the end.

At the mean-field level, we assume that \begin{enumerate*}[label=(\roman*)]
    \item the $\Phi$ fields are static, 
    \item the $W$ field can be neglected \footnote{As argued in previous studies, these turn out to vanish at the mean-field level anyway \cite{youDopingSpinorbitMott2012,seifertFractionalizedFermiLiquids2018b}}, and 
    \item that $\Phi$ is a diagonal matrix $\Phi_{\mu\nu} = \frac{1}{4} \delta_{\mu\nu} \Phi^{\mu}$.
\end{enumerate*}
The last assumption is a simplification which amounts to only having non-zero condensates of the form $\langle \iu \chi^{\mu}_{i}\chi^{\nu}_{j} \rangle$ for $\mu = \nu$.
In this scenario, the inverse of the interaction matrix $(V^{\mathsf{a}})^{-1}$ is simple to compute, since 
\begin{equation}
    (V^{\mathsf{a}})_{\mu\nu} = \frac{K}{16} M^{\mathsf{a}}_{\mu\nu} M^{\mathsf{a}}_{\mu\nu}= \frac{K}{16} \abs{M^{\mathsf{a}}}_{\mu\nu} \equiv \frac{K}{16} \abs{M^{\mathsf{a}}}^{-1}_{\mu\nu}.
\end{equation}

Furthermore, we assume that \begin{enumerate*}[label=(\roman*)]
    \setcounter{enumi}{3}
    \item $\Phi^{\mu}(\mathsf{a}) = \delta^{\mu0} u^{\mathsf{a}} + \delta^{\mu\mathsf{a}} u^{0}$, where the $u$'s are simply the mean-field values,
\end{enumerate*}
to connect with the mean-field form found by Ref.~\cite{youDopingSpinorbitMott2012}.
Invoking these assumptions, the mean-field action reads
\begin{equation}\label{eq:mf_action}
\begin{split}
    \s_{\mathrm{mf}} &= \frac{4 \beta N}{K} \sum_{\mathsf{a}=1}^{3} u^{\mathsf{a}}u^{0} + \frac{1}{2}\int_{0}^{\beta} \D\tau \sum_{i\in\Lambda} \chi^{\mu}_{i}\partial_{\tau} \chi^{\mu}_{i} \\
    &- \frac{1}{2} \int_{0}^{\beta} \D \tau \sum_{\mathsf{a}=1}^{3} \sum_{\langle i,j\rangle_{\mathsf{a}}} \iu u^{0} \chi^{\mathsf{a}}_{i} \chi^{\mathsf{a}}_{j} + \iu u^{\mathsf{a}} \chi^{0}_{i} \chi^{0}_{j}.
\end{split}
\end{equation}
Being quadratic in the Majorana fields $\chi$, the $\chi$'s can be integrated out exactly which in turn yields an effective mean-field free energy for the $u$'s.
Extremizing this free energy yields the following zero-temperature saddle-point equations 
\begin{subequations}\label{eq:saddle_point_eqns}
\begin{align}
    u^{\mathsf{a}} &= - \frac{1}{2}\frac{K}{4} \sign(u^0), \quad \and \\
    u^{0} &= - \frac{1}{6} \frac{K}{4} \sign(u^{\mathsf{a}}) \frac{1}{N} \sum_{\v{k} \in \varhexagon } \abs{\delta(\v{k})}, 
\end{align}
\end{subequations}
where $\delta(\v{k}) \coloneqq \sum_{\mathsf{a}} \exp(\iu\v{k}\cdot\v{n}_{\mathsf{a}})$, $\v{n}_{3} \coloneqq \v{0}$ and $\v{n}_{1,2} \coloneqq a(\sqrt{3},\pm 1)^{\mathsf{T}}/2$ are the lattice translation vectors of the hexagonal lattice, and $\varhexagon$ denotes the first Brillouin zone (consult Refs.~\cite{youDopingSpinorbitMott2012,seifertFractionalizedFermiLiquids2018b} for details).
Eqs.~\eqref{eq:saddle_point_eqns} coincide with those found in \cite{seifertFractionalizedFermiLiquids2018b} and upon scaling $K$ by $4$ with those in \cite{youDopingSpinorbitMott2012}.
As discussed by Ref.~\cite{seifertFractionalizedFermiLiquids2018b}, the discrepancy of the factor of $4$ is an artifact of the spin representation used, reflecting the fact that some degrees of freedom are gauge-equivalent upon explicitly enforcing $J^{\mathsf{a}} = 0$, while the connection between the results is established by the particular mean-field ansatz (assumption (\romannumeral 4 )).
As noted by Ref.~\cite{youDopingSpinorbitMott2012}, projecting this state onto the physical Hilbert space of isospin singlets yields the exact ground state constructed by Kitaev \cite{kitaevAnyonsExactlySolved2006}.
Since the choice of spin representation is qualitatively irrelevant, we will use the Kitaev representation in Eq.~\eqref{eq:kitaev_rep} henceforth.

Since $u^{\mu}$ are simply $\CC$-numbers, it is clear from the mean-field action in Eq.~\eqref{eq:mf_action} that the $\chi^{0}$ fields will have a graphene-like dispersion 
\begin{equation}\label{eq:chi0_dispersion}
    E_{\chi^{0}}(\v{k})= \abs{\sum_{\mathsf{a}=1}^{3} u^{\mathsf{a}} \e^{\iu\v{k}\cdot\v{n}_{\mathsf{a}}}},
\end{equation}
while the $\chi^{\mathsf{a}}$ modes are non-dispersive, with the gap given by $\abs{u^{0}}$. 
These dispersions are shown in Fig.~\ref{fig:low_energy_approx}.

\section{Low-energy effective theory}

\begin{figure}[htb]
	\centering
	\includegraphics[width=\columnwidth]{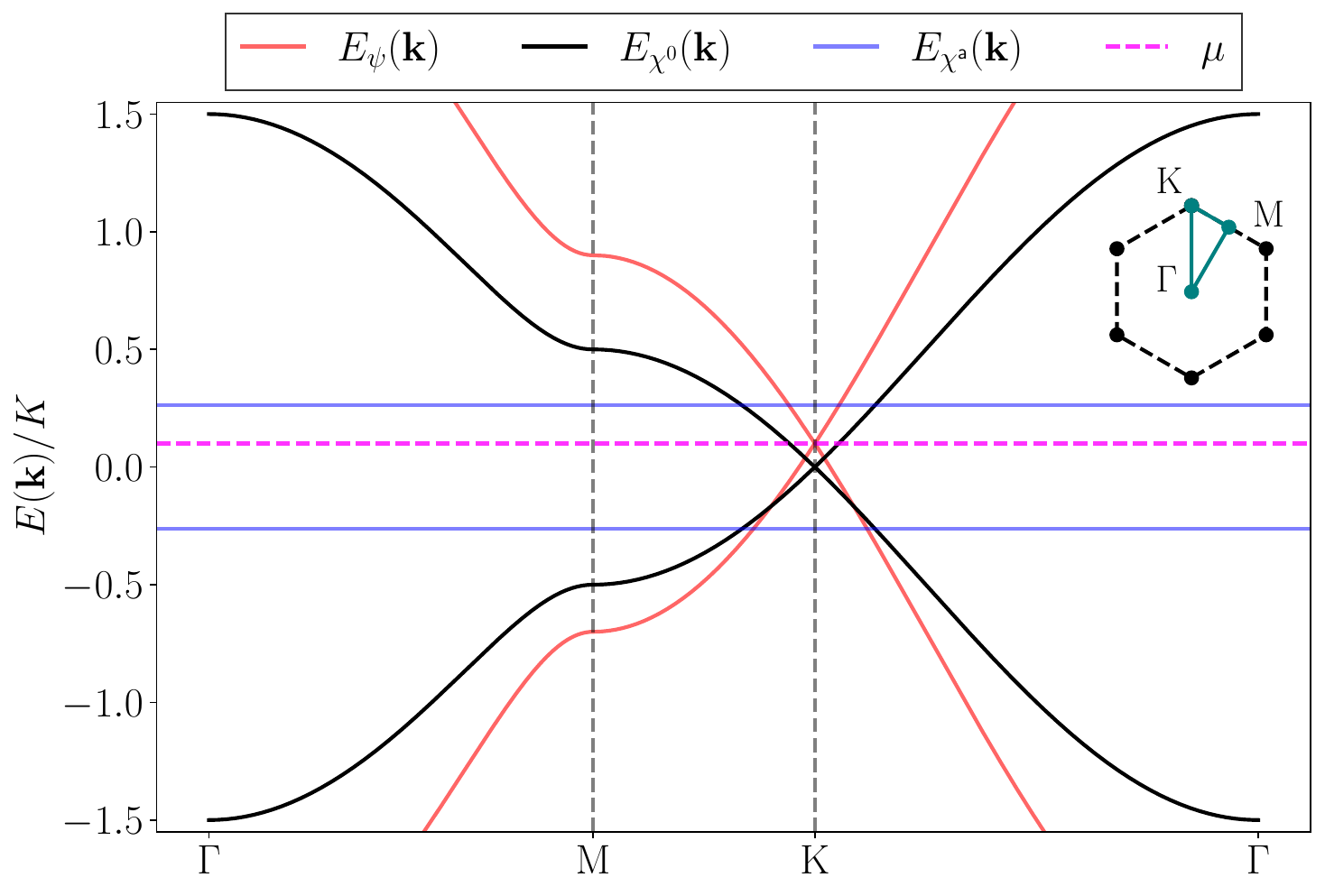}
	\caption{Dispersion of the Majorana fermions together with those of the conduction electrons and their chemical potential, similar to Fig.~2 of Ref.~\cite{choiTopologicalSuperconductivityKondoKitaev2018}. The path traversed in the Brillouin zone is illustrated in the inset as the teal line.
	}
	\label{fig:low_energy_approx}
\end{figure}

%\textcolor{red}{
In comparing the mean-field theory with Kitaev's exact solution, one identifies the $u^{\mathsf{a}}$ field as the $\ZZ_2$ gauge field.
Having energy gaps of order $K$, this field should be treated as static in the low-energy limit.
Under this assumption, Ref.~\cite{seifertFractionalizedFermiLiquids2018b} showed that the spin-spin interaction induced by the Kondo-coupled Fermi liquid simply renormalises the Kitaev interaction strength by a correction of order $J^2/K$. 
Beyond the static limit of the visons, the electrons induce an RKKY interaction in the spin sector \cite{koganRKKYInteractionGraphene2011}.
However, any long-range order effectuated by such a term is suppressed by the vanishing Majorana density of states \cite{seifertFractionalizedFermiLiquids2018b}.
The spin liquid state of the Kitaev model is, therefore, not destabilized for small $J$ \cite{senthilFractionalizedFermiLiquids2003,seifertFractionalizedFermiLiquids2018b}, and we can approximate the Kitaev model by its fermionic mean-field action when working to leading order in $J/K$.  
In the low-energy regime, this corresponds to three flavors of massive, non-dispersive fermions and one flavor of massless Dirac fermions, with momenta restricted to lie within a small range around $\v{k} = \v{K}$.
The low-energy-projected action of the conduction elections also gives rise to Dirac fermions, with two flavors corresponding to the two Dirac cones at $\v{k} = \pm \v{K}$. 
The low-energy restriction of the bands corresponds to focusing on the vicinity of the $\mathrm{K}$ point in Fig.~\ref{fig:low_energy_approx}.
Regarding the ratio $t/K$, we assume that $K$ originates with a mechanism similar to the one responsible for the usual ferromagnetic Heisenberg interaction, in which case it is natural to take $K < t$. 

\begin{widetext}
Using these simplifications, the low-energy effective action of the Kitaev model reads

\begin{align}\label{eq:chi_low_energy}
    \s_{\chi} + \s_{\Phi\chi} \simeq \int_{0}^{\beta} \D \tau \sum_{\abs{\v{k}} < \Lambda} \left\{ \chi^{0\dagger}_{\v{k}}(\tau) \left( \mathbf{1} \partial_{\tau} + c_{\chi} \epsilon^{ij} \sigma_{i} k_{j} \right) \chi^{0}_{\v{k}}(\tau) 
    + \sum_{\mathsf{a}=1}^{3}\chi^{\mathsf{a}\dagger}_{\v{k}}(\tau) \left( \mathbf{1}\partial_{\tau} + \sigma^{3} m c_{\chi}^{2} \right) \chi^{\mathsf{a}}_{\v{k}}(\tau) \right\},
\end{align}
\end{widetext}
where $\epsilon^{ij}$ is the antisymmetric symbol, $\mathbf{1}$ is the $2\times2$ unit matrix, $\{\sigma^{i}\}_{i=1}^{3}$ are the Pauli matrices, and $\Lambda$ is some momentum cutoff appropriate for the projection onto the low-energy sector of the theory. 
The constants appearing in this action are determined from the mean-field solution and are given by $
    c_{\chi} \equiv \sqrt{3}u^{\mathsf{a}} a/2$ and 
    $
    m c_{\chi}^{2} = -u^{0}
    $
    (see Appendix~\ref{app:low_energy} for details).
The low-energy fields appearing in this action are two-component spinor fields constructed from the two sublattice flavors around the Dirac point $\v{k}=\v{K}$ and are to be found in Appendix~\ref{app:low_energy}.

Eq.~\eqref{eq:chi_low_energy} implies that the bare Majorana propagators are given by
\begin{subequations}
	\begin{align}
		(\mathscr{D}^{0})^{-1}(k) &= \iu\omega_n \mathbf{1} - c_{\chi} \epsilon^{ij}\sigma_{i}k_{j} \notag \\
        &\Rightarrow \mathscr{D}^0(k) = \frac{\iu \omega_n \mathbf{1}+ c_{\chi} \epsilon^{ij}\sigma_{i}k_{j} }{(\iu\omega_n)^2 - c_{\chi}^2 \v{k}^2} \\
		(\mathscr{D}^{\mathsf{a}})^{-1}(k) &= \iu\omega_n \mathbf{1} - \sigma^{3} m c_{\chi}^2 \notag \\
  &\Rightarrow \mathscr{D}^{\mathsf{a}}(k) = \frac{\iu\omega_n \mathbf{1} + \sigma^3 m c_{\chi}^2}{(\iu\omega_n)^2 - (m c_{\chi}^2)^2}.
	\end{align} 
\end{subequations}

For the conduction electrons, we find a low-energy action similar to that of $\chi^{0}$, except that there are two flavors ($\alpha = 1,2$) of low-energy fields for the conduction electrons corresponding to excitations around $\v{k} = \pm\v{K}$, and they additionally carry a spin index ($\sigma = \uparrow,\downarrow$)
\begin{align}
    \s_{\psi} \simeq & \notag \\  
    \int_{0}^{\beta}\D \tau &\sum_{\abs{\v{k}} < \Lambda }\sum_{\alpha \sigma} \psi_{\sigma\v{k}}^{\alpha\dagger}(\tau) \left( \mathbf{1}\partial_{\tau} + c_{\psi} \epsilon^{ij}\sigma_{i}k_{j} \right) \psi^{\alpha\mathstrut}_{\sigma \v{k}}(\tau), 
\end{align}
where $c_{\psi} \equiv \sqrt{3} at/2$ is in general a different effective velocity than $c_{\chi}$, and the chemical potential is omitted for brevity. 

The remaining part of the low-energy theory is the Kondo interaction.
Since our strategy is to eventually integrate out the low-energy modes of the Kitaev spin liquid, it is necessary to express the interaction using these coordinates rather than the original fields. 
By denoting the composite operator representing the spin of an electron at sublattice $\lambda$ as $\v{s}_{\lambda} \coloneqq \psi^{\dagger}_{\lambda} \bm{\sigma} \psi^{\mathstrut}_{\lambda}$ (suppressing all additional labels and functional dependencies of $\psi$) we find that 
\begin{align}\label{eq:low_e_kondo}
    \s_{\chi\psi} &\simeq \frac{J}{N} \int_{0}^{\beta} \D \tau \sum_{\abs{\v{k}_1},\abs{\v{k}_2} < \Lambda } \sum_{\lambda = A,B} s^{\mathsf{a}}_{\lambda, \v{k}_1-\v{k}_2}(\tau) \notag \\
    &\times\bigg[ \chi^{0\dagger}_{\v{k}_1}(\tau) \iu\mathcal{M}_{\lambda}^{\mathstrut} \chi^{\mathsf{a}\mathstrut}_{\v{k}_2}(\tau) - \chi^{\mathsf{a}\dagger}_{\v{k}_1}(\tau) \iu\mathcal{M}_{\lambda}^{\dagger} \chi^{0\mathstrut}_{\v{k}_2}(\tau) \bigg], 
\end{align}
where $\mathcal{M}_{\lambda}$ are $2\times2$ matrices derived in Appendix~\ref{app:kondo_term}.   

\section{Effective theory of the conduction electrons}

Using the schematic notation $\chi \coloneqq (\chi^{0} \quad \chi^{\mathsf{a}})^{\mathsf{T}}$, the non-interacting part of the action can be written as 
\begin{subequations}
\begin{align}
    \s_{0} &= \s_{\psi} + \sum_{k} \chi^{\dagger}_{k} \left(-\mathscr{D}^{-1}(k)\right) \chi_{k}^{\mathstrut}, \quad \text{with} \notag \\
    \mathscr{D}^{-1}(k) &\coloneqq \begin{pmatrix} 
		\mathscr{D}^{-1}_{0}(k) & \\
		& \mathscr{D}^{-1}_{\mathsf{a}}(k)
	\end{pmatrix},
\end{align}
and the Kondo interaction as 
\begin{align}
    \s_{\chi\psi} &= \sum_{k_1,k_2} \chi_{k_1}^{\dagger} \mathscr{C}(k_1 - k_2) \chi_{k_2}^{\mathstrut} \quad \text{with} \notag \\
    \mathscr{C}(q) &\coloneqq \begin{pmatrix}
        & \mathscr{C}^{\mathsf{a}}(q) \\
        \mathscr{C}^{\mathsf{a}\dagger}(-q) 
    \end{pmatrix},
\end{align}
and 
\begin{equation}
    \mathscr{C}^{\mathsf{a}}(q) \equiv \frac{J}{\beta N} \sum_{\lambda = A,B} s_{\lambda}^{\mathsf{a}}(q) \iu \mathcal{M}_{\lambda}.
\end{equation}
\end{subequations}
Integrating out the low-energy fields $\chi$ yields
\begin{align}
    \s_{\mathrm{eff}}[\Psi^{\dagger},\Psi] &= \s_{\psi} - \tr\log\left( - \mathscr{D}^{-1} + \mathscr{C}\right).
\end{align}
We now expand the tracelog in the formula above to leading order in $J$, i.e., leading order in the interaction $\mathscr{C}$, and neglect the constant term representing the mean-field free energy of the Kitaev model $\tr\log(-\mathscr{D}^{-1})$.
This yields
\begin{align}
    \s_{\mathrm{eff}}[\Psi^{\dagger},\Psi] \simeq \s_{\psi} + \tr\left(\mathscr{DC}\right) + \frac{1}{2} \tr\left(\mathscr{DCDC}\right).
\end{align}
The first correction $\mathcal{O}(J)$ vanishes exactly since the matrix $\mathscr{D}$ is diagonal while $\mathscr{C}$ is antidiagonal.
Since each $\mathscr{C}$ is bilinear in conduction electron fields, the leading correction term represents a perturbatively induced quartic interaction of $\mathcal{O}(J^2)$.

\subsection{Induced quartic interaction}

Let us examine the second-order term in more detail. 
By resolving the operator trace in momentum space and the trace of the outermost matrix grading we find 
\begin{align}
    \frac{1}{2} \tr\left(\mathscr{DCDC}\right) & \notag \\
     =\tr_{\CC^2} \sum_{k,q} &\sum_{\mathsf{a}=1}^{3} \mathscr{D}^{0}(k) \mathscr{C}^{\mathsf{a}}(q) \mathscr{D}^{\mathsf{a}}(k-q) \mathscr{C}^{\mathsf{a}\dagger}(+q).
\end{align}
Moreover, using the form of the propagators together with the explicit form of the $\mathcal{M}_{\lambda}$ matrices we can resolve the remaining matrix trace as well (see Appendix~\ref{app:potential} for details) and be left with
\begin{align}
    \frac{1}{2}&\tr \left(\mathscr{DCDC}\right)  \notag \\
    =&\frac{J^2}{\left(\beta N\right)^2} \sum_{k,q} \sum_{\mathsf{a}=1}^{3} \sum_{\lambda = A,B} \mathscr{D}^{0}_{0}(k) s_{\lambda}^{\mathsf{a}}(q) \mathscr{D}_{0}^{\mathsf{a}}(k-q) s_{\lambda}^{\mathsf{a}}(-q) \notag \\
    =& \frac{1}{\beta V} \sum_{q} \sum_{\mathsf{a},\lambda}\Gamma^{\mathsf{a}}(q) s_{\lambda}^{\mathsf{a}}(q) s_{\lambda}^{\mathsf{a}}(-q), \label{eq:quartic_int}
\end{align}
where 
\begin{align}
    \Gamma^{\mathsf{a}}(\iu\omega_m,\v{q}) &\equiv \frac{J^2 a^2}{\beta N} \sum_{\abs{\v{k}} < \Lambda} \sum_{n\in\ZZ} \mathscr{D}^{0}_{0}(\v{k},n) \mathscr{D}^{\mathsf{a}}_ {0}(\v{k}-\v{q},n - m), \notag 
\end{align}
denotes the interaction potential and the $0$ subscript on the propagators refer to their Matsubara frequency components.
Due to the simple form of the propagators, $\Gamma$ is in fact independent of the spatial transferred momentum $\v{q}$.
Moreover, working in the low-temperature and static limits, $\Gamma$ can be approximated by a negative constant value: $\Gamma^{\mathsf{a}}(\iu\omega_m) \approx - \gamma $ (see Appendix \ref{app:potential}). 

Let us express the four-fermion interaction in terms of the low-energy excitations of the $\psi$ field.
There are two ``band"-flavors of these at each $\pm \v{K}$, which have dispersions $\xi_{\v{p}\pm} = \epsilon_{\v{p}\pm} - \mu = \pm c_{\psi} \abs{\v{p}} - \mu$ with $\v{p}$ being a \textit{small} momentum around $\pm \v{K}$.
Denote these fields by $\Psi_{s \sigma}^{\alpha}(\v{p})$, where $s = \pm$ designates whether the dispersion is $\pm c_{\psi}\abs{\v{p}}$, and $\alpha = 1,2$ designates whether it refers to the $+\v{K}$ or $-\v{K}$ symmetry point, and $\sigma$ its spin, i.e.,
\begin{equation}\label{eq:low_E_psi}
	c_{\psi}\begin{pmatrix}
		&  p_{y} + \iu p_x  \\
		 p_{y} - \iu p_x  & 
	\end{pmatrix}
	\Psi_{\pm \sigma}^{\alpha}(\v{p})
	= 
	\pm c_{\psi} \abs{\v{p}} \Psi_{\pm \sigma}^{\alpha}(\v{p}).
\end{equation} 
The bases for which the low-energy Hamiltonian of the conduction electrons take the form \eqref{eq:low_E_psi} are given by 
\begin{equation}\label{eq:psi_low_e}
	\psi^{1}_{\v{p}\sigma} \coloneqq \begin{pmatrix}
		\psi_{B \v{K} + \v{p}\sigma} \\
		\psi_{A \v{K} + \v{p}\sigma} 
	\end{pmatrix}
	\:\:\:
	\text{and}
	\:\:\:
	\psi^{2}_{\v{p}\sigma}  \coloneqq \begin{pmatrix}
		\psi_{A \v{p} - \v{K}\sigma} \\
		-\psi_{B \v{p} - \v{K}\sigma} 
	\end{pmatrix}.
\end{equation}
By diagonalizing the matrix, we find the eigenvectors $F_{s=\pm}$ for the two eigenvalues $\pm c_{\psi}\abs{\v{p}}$.
These are given by    
\begin{equation}
	F_{\pm} = \frac{1}{\sqrt{2}} \begin{pmatrix}\displaystyle
		\pm \frac{p_y + \iu p_x}{\abs{\v{p}}} & 1 
	\end{pmatrix}^{\mathsf{T}}.
\end{equation}
Defining
\begin{subequations}
\begin{align}
	F_{\pm}^{1} &\coloneqq  \frac{1}{\sqrt{2}} \begin{pmatrix}
		1 & \displaystyle \pm \frac{p_y + \iu p_x}{\abs{\v{p}}} 
	\end{pmatrix}^{\mathsf{T}} \quad \text{and} \\ 
	F_{\pm}^{2} &\coloneqq \frac{1}{\sqrt{2}} \begin{pmatrix}\displaystyle
		\pm \frac{p_y + \iu p_x}{\abs{\v{p}}} & -1 
	\end{pmatrix}^{\mathsf{T}}
\end{align}
\end{subequations}
allows us to relate the $\lambda$-sublattice Fourier mode to the low-energy modes by
\begin{subequations}\label{eq:transform_low_e}
    \begin{align}
	\psi_{\lambda \v{K} + \v{p}\sigma} &= \sum_{s = \pm} \bar{F}_{s\lambda}^{1} (\v{p}) \Psi_{s\sigma}^{1} (\v{p})\quad \text{and}  \\
    \psi_{\lambda \v{p} - \v{K} \sigma} &= \sum_{s = \pm} \bar{F}_{s\lambda}^{2}(\v{p}) \Psi_{s\sigma}^{2} (\v{p}).
    \end{align}
\end{subequations}

In terms of the sublattice fermions, the interaction reads 
\begin{align*}
    \s_{\mathrm{int}} = - \frac{\gamma}{\beta V} \sum_{k k' q} \sum_{\mathsf{a},\lambda} \bar{\psi}_{\lambda k+q \alpha} \psi_{\lambda k \beta} \bar{\psi}_{\lambda k'-q \gamma} \psi_{\lambda k' \delta} \sigma^{\mathsf{a}}_{\alpha\beta} \sigma^{\mathsf{a}}_{\gamma\delta}.
\end{align*}
We can express this interaction in terms of the low-energy modes by shifting $k \mapsto k + K$ and $k' \mapsto k' - K$, where this shift is understood to only act on the spatial momenta.
The combination of the transformation defined in Eq.~\eqref{eq:transform_low_e} and a positive-signature permutation of the Grassmann fields yields
\begin{widetext}
\begin{equation}
\begin{split}
    \s_{\mathrm{int}} = -\frac{\gamma}{\beta V} \sum_{kk'q}\sum_{\mathsf{a},\lambda} \sum_{s_1\dotsb s_4}&F^{1}_{s_1 \lambda}(\v{k}+\v{q}) F^{2}_{s_2 \lambda} (\v{k}'-\v{q}) \bar{F}^{2}_{s_3 \lambda} (\v{k}') \bar{F}^{1}_{s_4\lambda}(\v{k})\sigma^{\mathsf{a}}_{\alpha\beta} \sigma^{\mathsf{a}}_{\gamma\delta}  \\
    \times&\bar{\Psi}^{1}_{s_1 \alpha }(k+q) \bar{\Psi}^{2}_{s_2 \gamma} (k'-q) \Psi^{2}_{s_3\delta}(k') \Psi^{1}_{s_4 \beta }(k),
\end{split}
\end{equation}
\end{widetext}
where the remaining momentum summations are to be understood as the low-energy restricted ones in the vicinity of the Dirac points of  Fig.~\ref{fig:low_energy_approx}.
Let us now feed the model with some physically justified assumptions to simplify it. 
We consider
\begin{enumerate*}[label=(\roman*)]
    \item only zero center-of-mass-momentum Cooper-pairs, i.e., $(k+q) = -(k'-q)$. This assumption naturally eliminates one momentum summation. Moreover,
    \item we assume only pairing between low-energy modes of one and the same band, i.e., $s_1=s_2=s_3=s_4$. Without loss of generality, we may take $\mu > 0$, in which case the accessible low-energy modes are in the $s=+$ band \footnote{We comment on the case of $s=-$ at a later stage.}.  
\end{enumerate*}
With these simplifications, we can do the summation over $\lambda$ and be left with
\begin{align}\label{eq:Sint}
    \s_{\mathrm{int}} \simeq - \frac{\gamma}{\beta V} \sum_{k k'} & \sum_{\mathsf{a}} \bar{g}_{\v{k}} g_{\v{k}'} \sigma^{\mathsf{a}}_{\alpha\beta} \sigma^{\mathsf{a}}_{\gamma\delta} \\
    \times &\bar{\Psi}_{+\alpha}^{1}(k) \bar{\Psi}_{+\gamma}^{2}(-k) \Psi^{2}_{+\delta}(-k') \Psi^{1}_{+\beta}(k'), \notag
\end{align}
where $
    g_{\v{k}} \coloneqq (k_x + \iu k_y)/\abs{\v{k}},
$
and $\gamma$ has been rescaled by $1/2$.

By introducing the composite fermion fields $\mathcal{B}_{s,m}(k)$ representing a Cooper pair with spin quantum number $s$ and $S_{z}$ quantum number $m$ one finds that the interaction can be brought into the form (see Appendix~\ref{app:spin_structure})
\begin{align*}
    	\s_{\mathrm{int}}\left[\bar{\Psi},\Psi\right] \simeq -\frac{\gamma}{\beta V} &\sum_{k k'} \bar{g}_{\v{k}} g_{\v{k}'}  \notag \\
     \times\Bigg[\sum_{m = -1,0,1} & \mathcal{B}^{\dagger}_{1,m}(k)  \mathcal{B}_{1,m}^{\mathstrut}(k')  
      - 3 \mathcal{B}^{\dagger}_{0,0}(k) \mathcal{B}_{0,0}^{\mathstrut}(k') \Bigg].
\end{align*}
The interaction is repulsive in the singlet channel $(s=0)$.
Moreover, the factors $g_{\v{k}}$ appearing in the potential are odd in $\v{k}$, making them incompatible with a spin-singlet gap.
We therefore discard the singlet term in the following and consider
\begin{equation*}
	\s_{\mathrm{int}}\left[\bar{\Psi},\Psi\right] \simeq -\frac{\gamma}{\beta V} \sum_{k k'} \bar{g}_{\v{k}} g_{\v{k}'} \sum_{m = -1,0,1} \mathcal{B}^{\dagger}_{1,m}(k)  \mathcal{B}_{1,m}^{\mathstrut}(k').
\end{equation*}

In two spatial dimensions or less, long-wavelength phase fluctuations preclude long-range order at $T > 0$ \cite{Hohenberg, MerminWagner}.
The normal state is restored by a loss of phase stiffness via a mechanism not captured by mean-field theory, at a considerably lower temperature than the mean-field critical temperature we could estimate from the above theory \cite{kosterlitzOrderingMetastabilityPhase1973, NelsonKosterlitz, loktevSuppressionSuperconductingTransition2009}.
We therefore focus on classifying the possible superconducting states arising from this interaction at $T=0$. 

%We focus on classifying the possible superconducting states arising from this interaction at $T=0$, since long-wavelength phase fluctuations preclude long-range order at $T > 0$ \cite{Hohenberg, MerminWagner}, and the restoration of the normal state at $T > 0$ in principle is not captured by mean-field theory \cite{kosterlitzOrderingMetastabilityPhase1973,NelsonKosterlitz}.

\subsection{BCS mean-field theory}

The form of the quartic interaction derived in the preceding section naturally leads to the definition of chiral $p$-wave superconducting order parameters
\begin{equation}\label{eq:gaps}
    \begin{split}
        \Delta_{m}(k) &\coloneqq g_{\v{k}} \left\langle \mathcal{B}_{1,m}(k)\right\rangle \quad \text{and} \\
        \bar{\Delta}_{m}(k) &\coloneqq \bar{g}_{\v{k}} \left\langle \mathcal{B}^{\dagger}_{1,m}(k)\right\rangle,
    \end{split}
\end{equation}
where the objects inside the brackets of Eq.~\eqref{eq:gaps} should be interpreted as the operators on Fock space, which until now have been represented by Grassmann-valued fields.
Approximating the interaction vertex as frequency-independent permits us to define the momentum-independent gaps 
\begin{equation}
    \Delta_{m} \coloneqq \frac{\gamma}{\beta V} \sum_{k} \Delta_{m}(k).
\end{equation}
Since the propagators for the $\Psi$ fermions are spin-degenerate, and the interaction potentials for each of the spin triplets are the same, all the triplet superconducting gap amplitudes will also be degenerate at the mean-field level. 

Because the quartic interaction does not mix the different triplet order parameters, any coupling between them in the effective theory will only appear to fourth order in $\Delta$ when integrating out the $\sfPsi$ field.
In particular, there will be a ``Josephson" term at this order which involves the cosine of \textit{twice} the phase of the spin-polarized triplet gaps $\Delta_{\pm1}$ relative to the phase of the unpolarized one $\Delta_{0}$.
In interpreting the effective field theory of the superconducting order parameters as the free energy, and noticing that the Josephson term multiplies an overall \textit{positive} coefficient, the relative phases are fixed to take values $\pi/2$ or $3\pi/2$.
The $\ZZ_2$-redundancy of the ground state manifold reflects the spontaneous breaking of time-reversal symmetry in the chiral $p$-wave superconducting state \cite{ngBrokenTimereversalSymmetry2009,bojesenTimeReversalSymmetry2013,bojesenPhaseTransitionsAnomalous2014}.  

We define an $8$-component spinor $\sfPsi$ to set the stage for integrating out the fermions of the theory, and later recast our mean-field decoupled action in the form of a Bogoliubov--de-Gennes (BdG) Hamiltonian
\begin{equation}
	\sfPsi^{\dagger}(k)\coloneqq \begin{pmatrix}
		\bar{\Psi}_{\uparrow}(k) & \bar{\Psi}_{\downarrow}(k) &
		\Psi_{\downarrow}(-k) & \Psi_{\uparrow}(-k)
	\end{pmatrix},
\end{equation} 
where $\Psi_{\sigma}(k) \coloneqq \left( \Psi^{1}_{+\sigma}(k) \quad \Psi^{2}_{+\sigma}(k) \right)^{\mathsf{T}}$.
The basis has a particle-hole grading generated by the Pauli matrices $\rho^{\mu}$, a spin-$1/2$ grading generated by the Pauli matrices $\sigma^{\nu}$ and a ``valley" grading generated by the Pauli matrices $\tau^{\lambda}$. 
The particle-hole grading leads to a doubling of the kinetic terms and requires symmetrizing the terms involving the superconducting gap.
Introducing this spinor and symmetrizing the action accordingly yields
\begin{subequations}
\begin{equation}
	\s_{\mathrm{mf}} = \frac{\beta V}{\gamma} \sum_{m} \bar{\Delta}_{m} \Delta_{m} + \frac{1}{2} \sum_{k} \sfPsi^{\dagger}_{k} (-\mathscr{G}^{-1})(k) \sfPsi_{k}^{\mathstrut},
\end{equation}
with 
\begin{align}
	\mathscr{G}^{-1}(k) &= \left(\iu\omega_n \rho^{0} - \xi_{\v{k}} \rho^{3}\right) \otimes \sigma^{0} \otimes \tau^{0} \notag \\
	+ \bar{g}_{\v{k}} \rho^{+}& \otimes \left[ \Delta_{\uparrow\uparrow} \sigma^{+} + \Delta_{\downarrow\downarrow} \sigma^{-} + \Delta_{\uparrow\downarrow} \sigma^{0} \right] \otimes \tau^{1} \\
	+ g_{\v{k}} \rho^{-} &\otimes \left[ \bar{\Delta}_{\uparrow\uparrow} \sigma^{+} + \bar{\Delta}_{\downarrow\downarrow} \sigma^{-} + \bar{\Delta}_{\uparrow\downarrow} \sigma^{0} \right] \otimes \tau^{1}, \notag
\end{align} 
\end{subequations}
where we introduced the short-hand notation $2 \sigma^{\pm} \coloneqq \sigma^{1} \pm \iu \sigma^{2}$ (analogously for $\rho$). 

%\textcolor{red}{
Assuming all spin triplet gaps to be degenerate $\Delta_{m} \equiv \Delta$, and fixing a choice of the relative phases compatible with the analysis of the free energy of the system: $\e^{\iu\phi_0} = 1, \e^{\iu\phi_{1}} = \iu $ and $\e^{\iu\phi_{-1}} = \iu$, we can derive the BCS gap equation for this system \cite{bardeenTheorySuperconductivity1957a}.
This is done by first integrating out the Grassmann fields $\sfPsi$ and $\sfPsi^{\dagger}$ and subsequently minimising the resulting free energy functional with respect to $\bar{\Delta}$, yielding the saddle-point equation
\begin{align*}
    3 &\frac{\beta V}{\gamma} \Delta = \frac{1}{2} \tr\left( \mathscr{G} \frac{\partial \mathscr{G}^{-1}}{\partial \bar{\Delta}}\right) \\
    &= \frac{1}{2} \sum_{\v{k}}\sum_{n\in\ZZ} \tr_{\CC^8} \Big\{ \mathscr{G}(\v{k},n) g_{\v{k}} \left(\rho^{-} \otimes (\sigma^{0} - \iu \sigma^{1}) \otimes \tau^{1} \right) \Big\}.
\end{align*}
Inserting for the Green's function and resolving the trace yields the familiar BCS gap equation,
\begin{equation}
    1 = \frac{2 \gamma }{3} \frac{1}{V} \sum_{\v{k}} \frac{1}{\sqrt{\xi_{\v{k}}^2 + \vert \tilde{\Delta}\vert^2}  }  \tanh\left( \frac{\beta}{2}\sqrt{\xi_{\v{k}}^2 + \vert\tilde{\Delta} \vert ^2} \right),
\end{equation}
where $\tilde{\Delta} \coloneqq \sqrt{2} \Delta$.
With a linear fermionic dispersion, one cannot approximate the density of states at the Fermi level as in normal BCS theory.
Doing the remaining integral carefully, in this case, yields
\begin{equation}\label{eq:T=0gap}
    \abs{\Delta} \simeq \sqrt{2} \abs{\mu} \exp\left( - \frac{3 \pi c^2_{\psi}}{2 \gamma \abs{\mu}} \left[ 1 - \frac{2}{3\pi c^2_{\psi}} \gamma \omega_c \right] \right),
\end{equation}
demonstrating that a zero-temperature gap amplitude exists in the weak-coupling limit as long as $\mu \neq 0$ \cite{kopninBCSSuperconductivityDirac2008,xuDeterminationGapSolution2017}.
Let us also remark that the quantity in the square bracket above needs to be strictly positive for this equation to make sense. 
Indeed, in our perturbative regime $\gamma \omega_c /c_{\psi}^2 \sim (J/K)^2 (K/t)^2 \ll 1$.
%}

%\textcolor{red}{
Having established a non-trivial superconducting state at zero temperature, we now suggest to interpret $\s_{\mathrm{mf}}$ as a mean-field Hamiltonian of the low-energy fermions. 
In doing so, we drop the frequency-dependence and multiply by $\beta$ to get the BdG Hamiltonian 
\begin{subequations}
	\begin{equation}
		H = \frac{1}{2} \sum_{\v{k}} \sfPsi_{\v{k}}^{\dagger} \mathcal{H}_{\v{k}}^{\mathstrut} \sfPsi_{\v{k}}^{\mathstrut},
	\end{equation}
    where
    \begin{equation}
        \mathcal{H}_{\v{k}} \equiv \begin{pmatrix}
			H_0(\v{k}) & K_{\v{k}} \\
			K^{\dagger}_{\v{k}} & -H_{0}^{\mathsf{T}}(-\v{k})
		\end{pmatrix},
    \end{equation}
	with $ H_{0}(\v{k}) \coloneqq \xi_{\v{k}} \mathbf{1}_{4} = H_0(-\v{k})$, and 
	\begin{align}\label{eq:K_bdg}
		K_{\v{k}} &\coloneqq \bar{g}_{\v{k}} \Delta \left[ \e^{\iu\phi_{1}} \sigma^{+} + \e^{\iu\phi_{-1}} \sigma^{-} + \e^{\iu \phi_{0}} \sigma^0 \right] \otimes \tau^{1}
	\end{align}
	Here, $K_{\v{k}} = -K^{\mathsf{T}}_{-\v{k}}$ and $H_0^{\dagger}(\v{k}) = H_0(\v{k})$.
\end{subequations}

\subsection{Symmetry aspects of the mean-field theory}

By construction, the BdG Hamiltonian displays an explicit particle-hole symmetry through the fact that 
\begin{equation}
	\mathcal{C} \mathcal{H}_{\v{k}} \mathcal{C}^{-1} = - \mathcal{H}_{-\v{k}}, \quad \text{with} \quad \mathcal{C} \coloneqq \rho^{1} \otimes \sigma^{0} \otimes \tau^{0} \mathsf{K},
\end{equation} 
where $\mathsf{K}$ is the anti-unitary operator implementing complex conjugation, and the charge-conjugation operator satisfies $\mathcal{C}^2 = + 1$.
Exhibiting neither time-reversal nor chiral symmetry, the BdG Hamiltonian places the superconductor in class D of the tenfold classification \cite{altlandNonstandardSymmetryClasses1997,chiuClassificationTopologicalQuantum2016a}. 
In $d=2$, its (strong) topological character is revealed by an integer ($\ZZ$) topological invariant, the first Chern number, which will be computed in the next section.

\section{Topological response to a $\mathrm{U}(1)$ gauge field}

The topological invariant characterizing the superconducting state can be extracted as the coefficient controlling the topological response of the system to a $U(1)$ gauge field \cite{zhangChernSimonsLandau1992,qiTopologicalFieldTheory2008,volovikFractionalChargeSpin1989,yakovenkoChernSimonsTermsField1990}.
We minimally couple the low-energy fermions to a $\mathrm{U}(1)$ gauge field $A$ via the substitution 
	$k \to k - e A(q),$ 
where $e$ is the charge of the fermions and $q$ is a slowly varying momentum, and subsequently integrate out the fermions. 
To leading order in $A$ 
\begin{align}
	\mathscr{G}^{-1}(k - eA(q)) &\simeq \mathscr{G}^{-1}(k) - eA_{\mu}(q) \frac{\partial \mathscr{G}^{-1}(k)}{\partial k_{\mu}} \notag \\
	&\eqqcolon \mathscr{G}^{-1}(k) - \Sigma(k,q).
\end{align}
Integrating out the fermions yields an effective action in the form
\begin{equation}\label{eq:seff_gauge_field}
	\s_{\mathrm{eff}}[A] = \s_{0}[A] - \frac{1}{2} \tr\log\left( - \mathscr{G}^{-1} + \Sigma \right),
\end{equation}
where $\s_{0}[A]$ is the usual Maxwell action of the $\mathrm{U}(1)$ gauge field and the factor of $1/2$ multiplying the tracelog comes from the particle-hole doubling of the basis used to formulate the mean-field action \footnote{That is, $\sfPsi$ and $\sfPsi^{\dagger}$ are in fact only one independent field, made manifest through $
    \sfPsi(-k)^{\mathsf{T}} \rho^{1} \otimes \mathbf{1}_{4} = \sfPsi^{\dagger}(k)
$}.
By rescaling the gauge field according to $A \mapsto a \coloneqq e A/(\beta V)$, one finds that the effective action contains a \textit{Chern-Simons term} (see Appendix~\ref{app:topology} for details)
\begin{equation}
	\s_{\mathrm{eff}}[A] \supset \iu \frac{\mathsf{k}}{4\pi} \int \D^3 x \epsilon^{\mu\nu\rho} a_{\mu} \partial_{\nu} a_{\rho}. 
\end{equation}
The level of the Chern-Simons term, $\mathsf{k}$, is the first Chern number of the system \cite{qiTopologicalFieldTheory2008}. 
From the computation presented in Appendix~\ref{app:topology}, we find that it is given by $\mathsf{k} = N_3/2$, with 
\begin{widetext}
	\begin{equation}
		N_3 =  \frac{1}{24\pi^2} \epsilon^{\mu\nu\rho} \int \D^3 k  \tr_{\CC^8} \left[\mathscr{G}(k) \frac{\partial \mathscr{G}^{-1}(k)}{\partial k_{\mu}} \mathscr{G}(k) \frac{\partial \mathscr{G}^{-1}(k)}{\partial k_{\nu}} \mathscr{G}(k) \frac{\partial \mathscr{G}^{-1}(k)}{\partial k_{\rho}} \right],
	\end{equation}
\end{widetext}
in accordance with Ref.~\cite{volovikUniverseHeliumDroplet2009}.
Resolving the matrix trace and performing the remaining integral under the usual assumptions of BCS theory yields $\mathsf{k} \simeq 2 \times \sign{\mu}$.

Let us briefly interpret the topological invariant for this system.
At $T=0$ and chemical potential $\mu > 0$, the gap amplitude $\Delta$ is finite and the system enters a chiral topological superconducting phase characterized by Chern number $\mathsf{k} = 2$. 
As $\mu$ is lowered to $0$, there is no Fermi surface to support the formation of Cooper pairs, and consequently, the gap amplitude $\Delta$ vanishes. 
What is more, the Chern number at $\mu =0$ is zero, rendering the state topologically trivial. 
Lowering $\mu$ even further again gives rise to a topological superconductor, now characterized by $\mathsf{k} = -2$. 
The $T=0$ transition between states of distinct topological nature is a \textit{quantum topological phase transition}, directly connected to the closing and reopening of the gap of the low-energy fermionic excitations as $\mu$ is tuned through $0$. 

The non-zero value of the Chern number for the superconductor implies the existence of gapless Majorana fermions at the boundary \cite{chiuClassificationTopologicalQuantum2016a}.
In particular, since $\mathsf{k}=\pm 2$ the system hosts a pair of such fermions, which effectively combine into one massless Dirac fermion \cite{satoMirrorMajoranaZero2014}.
The presence of chiral, complex edge modes and the Chern-Simons response to a $\mathrm{U}(1)$ gauge field establishes a close analogy to the quantum Hall effect \cite{volovikQuantumHallChiral1992}. 
The application of such a system in topologically protected quantum computing, however, relies on having Majorana edge modes displaying non-abelian statistics \cite{nayakNonAbelianAnyonsTopological2008}.
There have been multiple efforts to address the problem of producing non-abelian anyons from such spinful superconductors
\cite{IvanovNonAbelian2001, KawakamiZeroEnergyModes2011, satoMirrorMajoranaZero2014,huangChiralMajoranaEdge2022} and particularly prove their relevance to topological quantum computing \cite{lianTopologicalQuantumComputation2018,hePlatformChiralMajorana2019}, but we leave these considerations in the current model for future work.

\section{Summary and Discussion}

We have presented a detailed derivation of the superconducting instability induced in the metal of the Kondo-Kitaev model to leading order in the Kondo coupling.
Starting from a low-energy treatment of the Kitaev honeycomb model, we obtained a description of it in terms of Dirac fermions, which we in turn integrated out to establish an effective theory of the conduction electrons.
To leading order in the Kondo coupling, we found an induced attractive interaction between pairs of electrons giving rise to a superconducting instability with triplet pairing and chiral $p$-wave symmetry. 
The limit of vanishing mean-field parameters of the Kitaev model appears innocuous in the sense that it leaves the quartic interaction potential finite.
However, the existence of non-zero values of these parameters is what allows us to characterize the excitations out of the ground state and to sensibly integrate them out of the theory, producing such an interaction. 
The coexisting QSL state is therefore an implicit requirement for the induced interaction. 

The attractive interaction in the triplet channel is attributed to the form of the interaction induced by the Kitaev spin liquid, while the chiral $p_x + \iu p_y$ structure comes from the particular wavefunctions describing the low-energy excitations of the conduction electrons on a honeycomb lattice.
The $p_x + \iu p_y$ structure has been identified before as a possible symmetry associated with the superconducting state of doped graphene \cite{uchoaSuperconductingStatesPure2007}. However, it has been far less trivial to pinpoint a pairing mechanism giving rise to it.
In contrast to phonons on the honeycomb lattice \cite{thingstadPhononmediatedSuperconductivityDoped2020}, the spin fluctuations out of the ground state of the Kondo-Kitaev model have dispersions with a node in Fourier space close to that of the conduction electrons, making it possible to realize superconductivity at relatively small dopings.    
Due to the chiral momentum structure of the gap, the superconducting state spontaneously breaks time-reversal symmetry. 
One could imagine this giving rise to an edge current, which in turn would yield an effective magnetic field and consequently alter the ground state of the Kitaev model. 
However, the current responsible for this magnetic field will be $\sim \abs{\Delta}^2$ so this is a sub-leading effect that can safely be neglected in our perturbative treatment.

By our analysis, the system is found to be a chiral topological superconductor of class D with first Chern number given by $2 \sign{\mu}$. 
At $\mu = 0$ we expect no superconducting state to emerge since there is no Fermi surface to support the superconducting instability.
It is therefore reassuring to find a vanishing Chern number at $\mu=0$.
Although the QSL state responsible for the interaction features topological order, the topological nature of the superconducting state has to be understood rather as a result of the induced attractive interaction in the triplet channel combined with the low-energy structure of the proximate graphene-like metal. 
Nevertheless, a non-zero value of the Kitaev order parameters was what enabled integrating out the Majoranas in the first place.
Together with the particular form of the induced interaction, this is a crucial feature allowing for TSC to form. 

\begin{acknowledgments}
    We thank Kristian Mæland and Jens Paaske for useful discussions.
	We acknowledge support from the Norwegian
	Research Council through Grant No. 262633, “Center of Excellence on Quantum Spintronics”, as well as Grant No. 323766.
\end{acknowledgments}

\appendix

\section{Details of the low-energy effective theory}\label{app:low_energy}

In this appendix, we provide some details on the derivation of the low-energy effective theory. 
Let us first consider the Majorana fields, and assume the order-parameter fields $u^{\mu}$ to take their mean-field values.
We introduce Fourier transforms according to 
\begin{equation}
    \chi_{\lambda i}^{\mu} = \frac{1}{\sqrt{N}} \sum_{\v{k} \in \varhexagon /2 } \left[ \e^{\iu\v{k}\cdot\v{r}_i} \chi^{\mu}_{{\textstyle{\mathstrut}}\lambda \v{k}} + \e^{-\iu\v{k} \cdot \v{r}_{i} } \bar{\chi}^{\mu}_{{\textstyle{\mathstrut}}\lambda \v{k}}  \right], 
\end{equation}
where we restrict the sum to run over \textit{half} of the Brillouin zone, permitting us to treat the Fourier components $\chi^{\mathstrut}_{\v{k}}$ and $\bar{\chi}^{\mathstrut}_{\v{k}} \equiv \chi^{\mathstrut}_{-\v{k}}$ as independent degrees of freedom \cite{colemanOddfrequencyPairingKondo1994}.
The two-component field $\chi^{0}_{\v{k}} \coloneqq \left( -\iu \chi^{0}_{A \v{k}} \quad \chi^{0}_{B \v{k}} \right)^{\mathsf{T}}$ is governed by the Hamiltonian 
\begin{equation}
    H_{\chi^{0}} = -\sum_{\v{k} \in \varhexagon/2} \chi^{0\dagger}_{\v{k}} \begin{pmatrix}
        & \gamma(\v{k}) \\
        \bar{\gamma}(\v{k}) & 
    \end{pmatrix} 
    \chi_{\v{k}}^{0\mathstrut},
\end{equation}
where $\gamma(\v{k}) \equiv u^{\mathsf{a}} \left(1 + \e^{\iu\v{k}\cdot\v{n}_1} + \e^{\iu\v{k}\cdot\v{n}_2}\right)$, giving rise to the dispersion in Eq.~\eqref{eq:chi0_dispersion}.
The dispersion has a node at $\v{K} = \frac{4\pi}{3a} \hat{y}$.
Being concerned with low-energy physics, we expand $\gamma(\v{k})$ around this node to leading order in $\abs{\v{p}}a$ 
\begin{equation}
    \gamma(\pm \v{K} + \v{p}) \simeq  \mp c_{\chi} \left( p_y \pm \iu p_x \right),
\end{equation}
where $c_{\chi} \equiv  \sqrt{3} a u^{\mathsf{a}}/2 $. 
Strictly speaking, $u^{\mathsf{a}}$ can take both positive and negative signs, as shown by the saddle-point Eqs.~\eqref{eq:saddle_point_eqns}. 
However, fixing the sign in these equations simply corresponds to fixing a gauge in the $\ZZ_2$ gauge theory of the $u^{\mathsf{a}}$ field.
We may therefore, without loss of generality, take $u^{\mathsf{a}} >0$ in the definition of $c_{\chi}$, in which case it faithfully represents the effective velocity of the Dirac fermions. 
The low-energy Hamiltonian reads
\begin{equation}
    H_{\chi^{0}} \simeq \sum_{\abs{\v{p}} < \Lambda } \chi^{0\dagger}_{{\textstyle\mathstrut}\v{K} + \v{p}} \left[ c_{\chi} (\bm{\sigma} \times \v{p}) \cdot \hat{z}  \right] \chi^{0}_{{\textstyle\mathstrut}\v{K} + \v{p}}.
\end{equation}
In the following, we will simply drop the reference to the $\v{K}$ momentum and denote these fields by $\chi^{0}_{\v{p}}$.

Likewise, the two-component field $X^{\mathsf{a}}_{\v{k}} \coloneqq \begin{pmatrix}
        \chi^{\mathsf{a}}_{A\v{k}} & \chi^{\mathsf{a}}_{B\v{k}}
    \end{pmatrix} ^{\mathsf{T}} $
is governed by the Hamiltonian
\begin{equation*}
    H_{\chi^{\mathsf{a}}} = \sum_{\v{k}\in\varhexagon/2}\sum_{\mathsf{a}} X^{\mathsf{a}\dagger}_{\v{k}}
    \begin{pmatrix}
        & - \iu u^{0}\e^{\iu \v{k} \cdot \v{n}_{\mathsf{a}} } \\
        \iu u^{0}\e^{-\iu \v{k} \cdot \v{n}_{\mathsf{a}} } & 
    \end{pmatrix}
    X^{\mathsf{a}}_{\v{k}}.
\end{equation*}
By diagonalizing this Hamiltonian and again restricting to small momenta around $\v{K}$ we find that the low-energy edition of it is given by 
\begin{equation}
    H_{\chi^{\mathsf{a}}} \simeq \sum_{\abs{\v{p}} < \Lambda } \sum_{\mathsf{a}} \chi^{\mathsf{a}\dagger}_{{\textstyle\mathstrut}\v{K} + \v{p}} \begin{pmatrix}
        mc_{\chi}^2 & \\
        & - m c_{\chi}^2 
    \end{pmatrix}
    \chi^{\mathsf{a}}_{{\textstyle\mathstrut}\v{K} + \v{p}},
\end{equation}
where the two-component fields $\chi^{\mathsf{a}}$ are defined as 
\begin{equation}
    \chi^{\mathsf{a}}_{\v{k}} \coloneqq \frac{1}{\sqrt{2}} \begin{pmatrix}
        +\iu\e^{\iu \v{k} \cdot \v{n}_{\mathsf{a}}} \chi_{{\textstyle\mathstrut} A \v{k}} + \chi_{{\textstyle\mathstrut} B \v{k}}  \\
        -\iu\e^{\iu \v{k} \cdot \v{n}_{\mathsf{a}}} \chi_{{\textstyle\mathstrut} A \v{k}}+ \chi_{{\textstyle\mathstrut} B \v{k}} 
    \end{pmatrix},
\end{equation}
and $mc_{\chi}^2 \equiv -u^{0} > 0$.
As before, we restrict to negative $u^{0}$ (corresponding to positive $u^{\mathsf{a}}$) by fixing the gauge.
In the following, drop the reference to the $\v{K}$ momentum in the fields as above.

Analogous to the low-energy treatment of $\chi^{0}$, the low-energy-projected action of the conduction elections also gives rise to Dirac fermions, but in this case, two flavors corresponding to the two Dirac cones at $\v{k} = \pm \v{K}$ appear. 
It is straightforward to verify that the two-component fields 
\begin{equation}
    \psi^{1}_{\v{p}} \coloneqq \begin{pmatrix}
        \psi_{{\textstyle\mathstrut}B \v{K} + \v{p}} \\
        \psi_{{\textstyle\mathstrut}A \v{K} + \v{p}}
    \end{pmatrix}
    \quad \text{and} \quad
    \psi^{2}_{\v{p}} \coloneqq \begin{pmatrix}
        \psi_{{\textstyle\mathstrut}A \v{p} - \v{K}} \\
        -\psi_{{\textstyle\mathstrut}B \v{p} - \v{K}}
    \end{pmatrix},
\end{equation}
are governed by the low-energy Hamiltonian 
\begin{equation}
    H_{\psi} \simeq \sum_{\abs{\v{p}} < \Lambda} \sum_{\alpha \sigma} \psi^{\alpha\dagger}_{\sigma \v{p}} \left[ c_{\psi} (\bm{\sigma} \times \v{p}) \cdot \hat{z}\right] \psi^{\alpha\mathstrut}_{\sigma\v{p}},
\end{equation}
with $c_{\psi} \equiv \sqrt{3} a t / 2$.

\onecolumngrid

\section{The Kondo interaction in terms of low-energy excitations}\label{app:kondo_term}

Recall that the Kondo interaction is \textit{local} in real space, and therefore also local in the sublattice indices.
We would like to re-express it rather in terms of the components of the low-energy fields $\chi^{0}$ and $\chi^{\mathsf{a}}$.
In terms of Fourier components, the Kondo interaction reads
\begin{equation*}
    H_{J} \simeq \frac{J}{N} \sum_{\abs{\v{k}_1}, \abs{\v{k}_{2}} < \Lambda } \sum_{\lambda=A,B} \left[s^{\mathsf{a}}_{\lambda}(\v{k}_1 - \v{k}_2) \chi^{0\dagger}_{{\textstyle\mathstrut}\lambda\v{k}_1} \chi^{\mathsf{a}}_{{\textstyle\mathstrut}\lambda \v{k}_2} + \hc \right].    
\end{equation*}
Now, note that 
\begin{subequations}
\begin{align}
    \bar{\chi}^{0}_{A}\chi^{\mathsf{a}}_{A} &= \chi^{0\dagger} \mathcal{M}_{A}^{\mathstrut} \chi^{\mathsf{a}} \quad \text{with} \quad M^{\mathstrut}_{A} \coloneqq \frac{1}{\sqrt{2}} \begin{pmatrix}
        -\e^{-\iu\phi_{\mathsf{a}}} & \e^{-\iu\phi_{\mathsf{a}}} \\
        0 & 0
    \end{pmatrix}, \quad \text{and} \\
    \bar{\chi}^{0}_{B}\chi^{\mathsf{a}}_{B} &= \chi^{0\dagger} \mathcal{M}_{B}^{\mathstrut} \chi^{\mathsf{a}} \quad \text{with} \quad M^{\mathstrut}_{B} \coloneqq \frac{1}{\sqrt{2}} \begin{pmatrix}
        0 & 0\\
        1 & 1
    \end{pmatrix},
\end{align}
\end{subequations}
where $\phi_{\mathsf{a}} \simeq \v{K}\cdot \v{n}_{\mathsf{a}}$, and the two-component fields appearing on the right-hand side of the equations above refer to the low-energy fields identified in the previous section.
This establishes the form employed in Eq.~\eqref{eq:low_e_kondo}.

\section{Quartic interaction}

In this appendix, we elaborate on the intermediate steps taking us from the induced quartic interaction to the one formulated in terms of composite fermionic fields.
We first consider the interaction potential arising from the tracelog and next consider the spin-structure of this interaction.

\subsection{Interaction potential}\label{app:potential}
When computing the perturbatively induced quartic interaction, we need to compute the trace over the matrices appearing in the propagators of the low-energy Majorana fields as well as those appearing in $\mathscr{C}$.
Specifically, the trace reads
    \begin{align}
	\tr_{\CC^2}\sum_{kq} &\sum_{\mathsf{a}} \mathscr{D}^{0}(k) \mathscr{C}^{\mathsf{a}}(q) \mathscr{D}^{\mathsf{a}}(k-q) \mathscr{C}^{\mathsf{a}\dagger}(+q) \notag \\
    &= \frac{J^2}{\left(\beta N\right)^2} \sum_{kq} \sum_{\mathsf{a}} \tr_{\CC^2}\bigg\{ \mathscr{D}^{0}(k) \Big[ s_{A}^{\mathsf{a}} (q) \iu\mathcal{M}_A + \iu s_{B}^{\mathsf{a}}(q)  \mathcal{M}_B \Big] 
	 \mathscr{D}^{\mathsf{a}}(k-q) \left[ - s_{A}^{\mathsf{a}}(-q) \iu  \mathcal{M}_{A}^{\dagger} - s_{B}^{\mathsf{a}}(-q) \iu \mathcal{M}_{B}^{\dagger} \right] \bigg\}.
\end{align}
At this point, a key observation to be made is that all terms except those coming from the frequency part of \textit{both} propagators will yield a vanishing result as they will give rise to either an integral over an odd function or a Matsubara sum of an odd summand. 
The only relevant matrix trace we have to perform is therefore 
\begin{equation}\label{eq:traces_1}
	\tr_{\CC^2}\left( \sigma^{\mu} \mathcal{M}_{i} \mathbf{1} \mathcal{M}_{j}^{\dagger} \right) = \delta_{ij} \delta_{\mu0},
\end{equation}
which yields the form of the interaction shown in Eq.~\eqref{eq:quartic_int} with 
\begin{align}
	\Gamma^{\mathsf{a}}(\iu\omega_m,\v{q}) &\equiv + \frac{J^2 \mathfrak{v}}{\beta N} \sum_{\abs{\v{k}} < \Lambda} \sum_{n\in\ZZ} \mathscr{D}^{0}_{0}(\v{k},n) \mathscr{D}^{\mathsf{a}}_ {0}(\v{k}-\v{q},n - m) \notag\\
    &= -\frac{\mathcal{J}^2}{\beta}\sum_{n\in\ZZ} \int\limits_{\abs{\v{k}} < \Lambda} \frac{\D^2 k}{(2\pi)^2} \frac{\omega_n(\omega_n - \omega_m)}{(\omega_n^2 + c_{\chi}^2 \v{k}^2) \left( (\omega_n - \omega_m)^2 + (mc_{\chi}^2)^2\right)} \notag \\
    &= - \frac{\mathcal{J}^2 }{4\pi c_{\chi}^2} \frac{1}{\beta} \sum_{n\in\ZZ} \frac{\omega_{n}(\omega_n - \omega_m)}{(\omega_n-\omega_m)^2 + (m c_{\chi}^2)^2} \log\left( 1 + \frac{c_{\chi}^2 \Lambda^2}{\omega_{n}^2}\right). \label{eq:Gamma_expr}
\end{align}
where we have absorbed two factors of $a^2 = V/N \eqqcolon \mathfrak{v}$ into the new coupling constant $\mathcal{J}$.
\begin{figure}[htb]
	\centering
	\includegraphics[width=0.65\columnwidth]{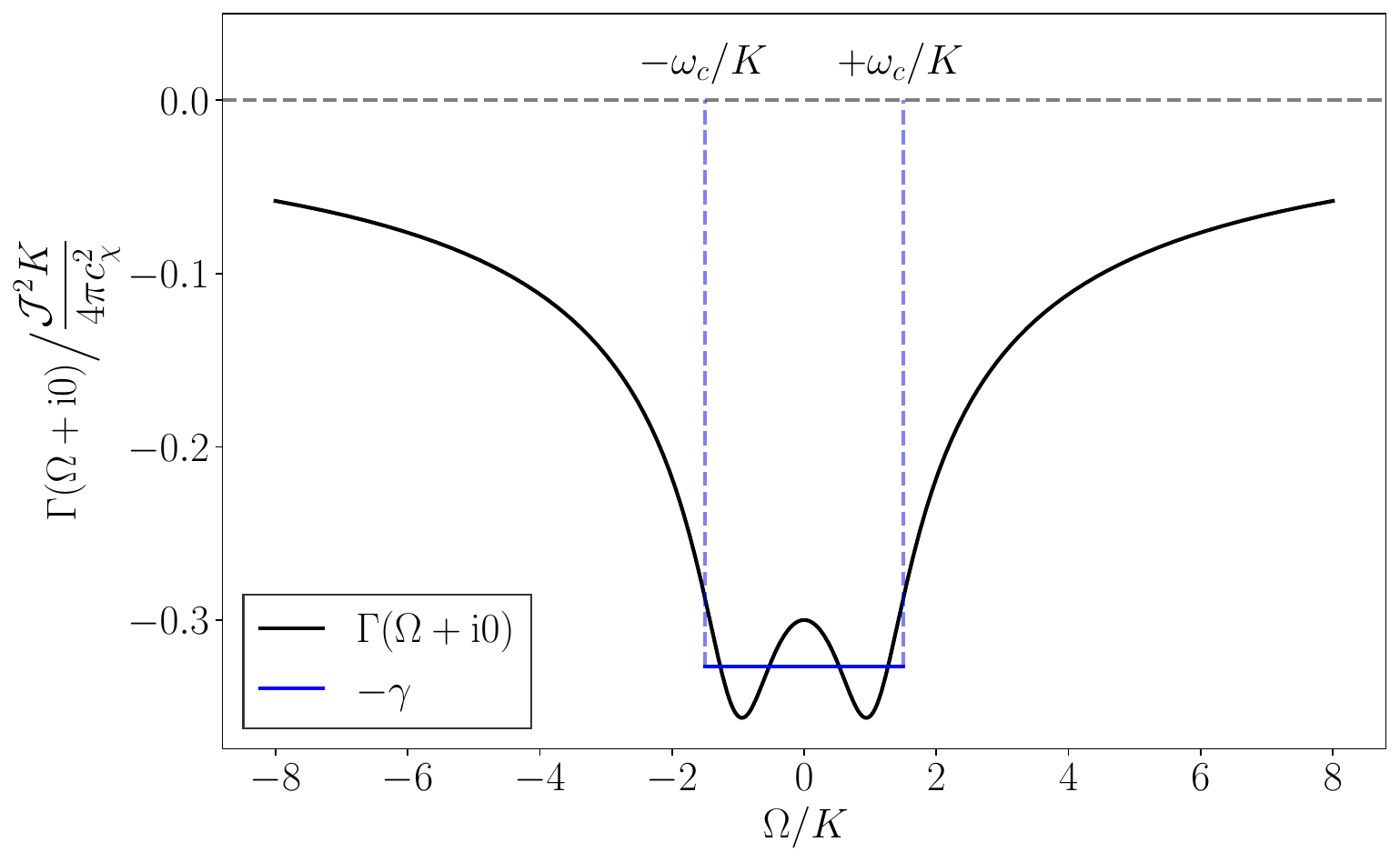}
	\caption{Zero-temperature interaction vertex as a function of real frequency, together with its constant approximation.}
	\label{fig:gamma_approx}
\end{figure}

By the analytic continuation of the argument $\iu\omega_m \to \Omega + \iu 0$, we can interpret $\Omega$ as the energy transfer of the two-body interaction defined by the quartic term. 
%\textcolor{red}{
Since both $mc_{\chi}^2$ and $c_{\chi}\Lambda$ are of order $K$, $K$ provides a natural energy scale for the remaining Matsubara sum.
Putting both of these equal to $K$ permits us to write
\begin{equation}
    \Gamma^{\mathsf{a}}(\iu\omega_{m}) = - \frac{\mathcal{J}^2 K }{4\pi c_{\chi}^2} F(\iu\omega_{m}/K),
\end{equation}
where we evaluate $F(\Omega/K + \iu 0)$ numerically in the zero-temperature limit. 
In the spirit of BCS theory, we approximate $F(\Omega/K + \iu 0)$ by its average value on a finite frequency interval.
This endows the mean-field theory with an energy cutoff, $\omega_c$, which in this case is chosen to be the bandwidth of the Majorana dispersion in Fig.~\ref{fig:low_energy_approx}.
The rationale for this is that $\Omega$ also corresponds to the energy carried by the virtual pair of Majoranas exchanged by the physical electron pairs. 
Extrapolating $\Gamma(\Omega + \iu 0)$ to $\Omega$ larger than approximately the bandwidth of the Majoranas is therefore unphysical.
This is illustrated in Fig.~\ref{fig:gamma_approx}.
This justifies working in the static limit and approximating the vertex by some representative negative constant: $\Gamma \approx - \gamma = \mathrm{const}$. 
%}

\subsection{Spin structure}\label{app:spin_structure}

Now, consider the interaction appearing in Eq.~\eqref{eq:Sint}.
By using the identity
\begin{equation}
    \sum_{\mathsf{a}=1}^{3} \sigma^{\mathsf{a}}_{\alpha\beta} \sigma^{\mathsf{a}}_{\gamma\delta} = 2 \delta_{\alpha\delta} \delta_{\beta\gamma} - \delta_{\alpha\beta} \delta_{\gamma\delta},
\end{equation}
and splitting the remaining spin sums into the terms where the two spins are equal and opposite respectively, we can write the induced interaction in Eq.~\eqref{eq:Sint} as 
\begin{alignat}{2}
	\s_{\mathrm{int}}\left[\bar{\Psi},\Psi\right] &= -\frac{\gamma}{\beta V} \sum_{k k'}\sum_{\alpha \in \{\uparrow,\downarrow\}} \bar{g}_{\v{k}} g_{\v{k}'} \bigg[ \bar{\Psi}^{1}_{+\alpha}(k)\bar{\Psi}^{2}_{+\alpha}(-k) \Psi^{2}_{+\alpha} (-k')\Psi^{1}_{+\alpha}(k') &+2\bar{\Psi}^{1}_{+\alpha}(k) \bar{\Psi}^{2}_{+\bar{\alpha}}(-k) \Psi^{2}_{+\alpha}(-k') \Psi^{1}_{+\bar{\alpha}}(k') \notag \\
	& &-\bar{\Psi}^{1}_{+\alpha}(k) \bar{\Psi}^{2}_{+\bar{\alpha}}(-k) \Psi^{2}_{+\bar{\alpha}}(-k') \Psi^{1}_{+\alpha}(k') \bigg] \notag  \\
	&\equiv -\frac{\gamma}{\beta V} \sum_{k k'} \bar{g}_{\v{k}} g_{\v{k}'} \Big[ A(k,k') + 2 B(k,k') - C(k,k') \Big], &
\end{alignat}
where the definitions of $A,B$, and $C$ will turn out to be useful in a moment.
In the above, we used the notation $\bar{\uparrow} \equiv \downarrow$ and $\bar{\downarrow} \equiv \uparrow$.
Now, let us introduce the following composite fermion fields 
\begin{subequations}
	\begin{align}
		&\mathcal{B}_{1,1}(k) \coloneqq \Psi^{2}_{+\uparrow}(-k) \Psi^{1}_{+\uparrow}(k)  &&\mathcal{B}_{1,-1}(k) \coloneqq \Psi^{2}_{+\downarrow}(-k) \Psi^{1}_{+\downarrow}(k) \\
		&\mathcal{B}_{1,0}(k) \coloneqq \frac{1}{\sqrt{2}}\Bigg[ \Psi^{2}_{+\uparrow}(-k) \Psi^{1}_{+\downarrow}(k) + \Psi^{2}_{+\downarrow}(-k) \Psi^{1}_{+\uparrow}(k) \Bigg]  &&\mathcal{B}_{0,0}(k) \coloneqq \frac{1}{\sqrt{2}}\Bigg[ \Psi^{2}_{+\uparrow}(-k) \Psi^{1}_{+\downarrow}(k) - \Psi^{2}_{+\downarrow}(-k) \Psi^{1}_{+\uparrow}(k) \Bigg].	\end{align}
\end{subequations}
That is, $\mathcal{B}_{s,m}$ is the Cooper pair with spin quantum number $s$ and $S_z$ quantum number $m$.
Now, notice that (suppressing momentum dependence for brevity)
\begin{equation}
	A = \mathcal{B}_{1,1}^{\dagger} \mathcal{B}_{1,1}^{\mathstrut} +\mathcal{B}_{1,-1}^{\dagger} \mathcal{B}_{1,-1}^{\mathstrut}, \quad  \mathcal{B}_{0,0}^{\dagger} \mathcal{B}_{0,0}^{\mathstrut} = \frac{1}{2} \left( C - B \right), \quad \text{and}\quad \mathcal{B}_{1,0}^{\dagger} \mathcal{B}_{1,0}^{\mathstrut} = \frac{1}{2} \left( C + B \right),
\end{equation}
so that 
\begin{equation}
	C = \mathcal{B}_{1,0}^{\dagger} \mathcal{B}_{1,0}^{\mathstrut} + \mathcal{B}_{0,0}^{\dagger} \mathcal{B}_{0,0}^{\mathstrut} \quad \text{and} \quad B = \mathcal{B}_{1,0}^{\dagger} \mathcal{B}_{1,0}^{\mathstrut} - \mathcal{B}_{0,0}^{\dagger} \mathcal{B}_{0,0}^{\mathstrut}.
\end{equation}
Thus, 
\begin{align}
	2 B - C &= 2 \left( \mathcal{B}_{1,0}^{\dagger} \mathcal{B}_{1,0}^{\mathstrut} - \mathcal{B}_{0,0}^{\dagger} \mathcal{B}_{0,0}^{\mathstrut} \right) - \left( \mathcal{B}_{1,0}^{\dagger} \mathcal{B}_{1,0}^{\mathstrut} + \mathcal{B}_{0,0}^{\dagger} \mathcal{B}_{0,0}^{\mathstrut} \right) = \mathcal{B}_{1,0}^{\dagger} \mathcal{B}_{1,0}^{\mathstrut} - 3 \mathcal{B}_{0,0}^{\dagger} \mathcal{B}_{0,0}^{\mathstrut}.
\end{align}
Using these results, we can write the interaction as 
\begin{equation*}
	\s_{\mathrm{int}}\left[\bar{\Psi},\Psi\right] = -\frac{\gamma}{\beta V} \sum_{k k'} \bar{g}_{\v{k}} g_{\v{k}'} \left[ \sum_{m = -1,0,1} \mathcal{B}^{\dagger}_{1,m}(k)  \mathcal{B}_{1,m}^{\mathstrut}(k') - 3 \mathcal{B}^{\dagger}_{0,0}(k) \mathcal{B}_{0,0}^{\mathstrut}(k')\right],
\end{equation*}
as advertised in the main text.

\section{Topological invariant}\label{app:topology}

In this appendix, we elaborate on the derivation of the topological invariant. 
Expanding the tracelog of Eq.~\eqref{eq:seff_gauge_field} to second order in the gauge field and resolving the operator trace in momentum space yields 
\begin{align}
	\s_{\mathrm{eff}}[A] &\supset +\frac{1}{4} \tr \left(  \mathscr{G}\Sigma\mathscr{G}\Sigma \right) \notag \\
	&= \frac{1}{4}\sum_{kq} \tr_{\CC^8} \left[ \mathscr{G}(k) \Sigma(k,q) \mathscr{G}(k-q) \Sigma(k,-q)\right] \notag \\
	&= \frac{e^2}{4}\sum_{kq} \tr_{\CC^8} \left[ \mathscr{G}(k) A_{\mu}(q) \frac{\partial \mathscr{G}^{-1}(k)}{\partial k_{\mu}} \mathscr{G}(k-q) A_{\nu}(-q) \frac{\partial \mathscr{G}^{-1}(k)}{\partial k_{\nu}} \right]. 
\end{align}
To simplify further, rescale the field according to $A \mapsto a \coloneqq e A/(\beta V)$ where $\beta V$ is the ``volume" of the system, and expand 
\begin{equation}
	\mathscr{G}(k-q) \approx \mathscr{G}(k) - q_{\rho} \frac{\partial \mathscr{G}(k)}{\partial k_{\rho}} = \mathscr{G}(k) + q_{\rho} \mathscr{G}(k) \frac{\partial \mathscr{G}^{-1}(k)}{\partial k_{\rho}} \mathscr{G}(k),
\end{equation}
where the last equality follows from the fact that 
\begin{equation}
	0 = \partial_{k} \left( \mathscr{GG}^{-1} \right) = (\partial_{k} \mathscr{G})\mathscr{G}^{-1} + \mathscr{G}(\partial_{k} \mathscr{G}^{-1}).
\end{equation}
Inserting this into the quadratic term in the $a$ fields, and focusing on the contribution from the term linear in $q$ we find 
\begin{equation}
	\s_{\mathrm{eff}}[A] \supset \frac{1}{4\pi} \int \frac{\D^3 q}{(2\pi)^3} a_{\mu}(q) q_{\rho} a_{\nu}(-q) \mathsf{C}^{\mu\rho\nu}
\end{equation}
with
\begin{equation}
	\mathsf{C}^{\mu\rho\nu} \equiv \pi \int \frac{\D^3 k}{(2\pi)^3} \tr_{\CC^8} \left[\mathscr{G}(k) \frac{\partial \mathscr{G}^{-1}(k)}{\partial k_{\mu}} \mathscr{G}(k) \frac{\partial \mathscr{G}^{-1}(k)}{\partial k_{\rho}} \mathscr{G}(k) \frac{\partial \mathscr{G}^{-1}(k)}{\partial k_{\nu}} \right].
\end{equation}

Now, take notice of the following. 
The integral of $a_{\mu} q_{\rho} a_{\nu}$ vanishes unless the coefficient $\mathsf{C}$ is antisymmetric in $\mu$ and $\nu$, which can be seen easily by going to real-space and doing integration by parts. 
Moreover, the trace term that multiplies it is cyclic in all indices, meaning that we can extend the antisymmetry to any pair of indices. 
Hence we can write $\mathsf{C}^{\mu\rho\nu} = \epsilon^{\mu\rho\nu} \mathsf{k}$ and by contracting the expression with the Levi-Civita symbol $\epsilon^{\mu\rho\nu}$ we find
\begin{align}
	3! \mathsf{k} &= \pi \epsilon^{\mu\rho\nu}\int \frac{\D^3 k}{(2\pi)^3} \tr_{\CC^8} \left[\mathscr{G}(k) \frac{\partial \mathscr{G}^{-1}(k)}{\partial k_{\mu}} \mathscr{G}(k) \frac{\partial \mathscr{G}^{-1}(k)}{\partial k_{\rho}} \mathscr{G}(k) \frac{\partial \mathscr{G}^{-1}(k)}{\partial k_{\nu}} \right] \notag \\
	&\Leftrightarrow  \mathsf{k} = \frac{1}{2} \frac{1}{24\pi^2} \epsilon^{\mu\rho\nu} \int \D^3 k  \tr_{\CC^8} \left[\mathscr{G}(k) \frac{\partial \mathscr{G}^{-1}(k)}{\partial k_{\mu}} \mathscr{G}(k) \frac{\partial \mathscr{G}^{-1}(k)}{\partial k_{\rho}} \mathscr{G}(k) \frac{\partial \mathscr{G}^{-1}(k)}{\partial k_{\nu}} \right]. \label{eq:CS_level}
\end{align} 
Hence, the term we have computed corresponds to a Chern-Simons term 
\begin{equation}
	\s_{\mathrm{eff}}[a] \supset \iu\frac{\mathsf{k}}{4\pi} \int \D^3 x  \epsilon^{\mu\nu\rho} a_{\mu} \partial_{\nu} a_{\rho}, 
\end{equation}
with the level given by $\mathsf{k}$ in Eq.~\eqref{eq:CS_level}.

Due to the antisymmetry of $\mathsf{C}^{\mu\rho\nu}$, it suffices to compute it for $\mu=0,\rho=1$ and $\nu=2$, and multiplying by $3!$ to obtain $\mathsf{k}$.
Moreover, since $\partial_{k_{0}} \mathscr{G}^{-1} = \iu\mathbf{1}_{8}$, we find 
 \begin{equation}
 	\mathsf{k} = \frac{\iu}{8\pi^2} \int \D^2 k \int \D k_{0} \tr_{\CC^8} \Big[ \mathscr{G} \mathscr{G}\left( \partial_{k_{1}} \mathscr{G}^{-1}  \right) \mathscr{G}\left( \partial_{k_{2}} \mathscr{G}^{-1} \right) \Big].
 \end{equation}
By expressing the Hamiltonian as $\mathcal{H}_{\v{k}} = \v{m}_{\v{k}} \cdot \v{U}$, where $\v{U}$ is a vector of $8\times8$ matrices, we find 
\begin{align}
	\mathsf{k} = \frac{\iu}{8\pi^2}\int \D^2 k \int \D k_{0}  \tr_{\CC^8} \left\{ \left[\left( \iu k_0 \mathbf{1} - \v{m}_{\v{k}} \cdot\v{U} \right)^{-1} \right]^2 U^{\alpha} \left( \iu k_0 \mathbf{1} - \v{m}_{\v{k}} \cdot \v{U} \right)^{-1} U^{\beta} \right\} \partial_{k_{1}} m^{\alpha}_{\v{k}}   \partial_{k_{2}} m^{\beta}_{\v{k}} . \label{eq:N3}
\end{align} 
Now, we fix a choice of relative phases of the BdG Hamiltonian in Eq.~\eqref{eq:K_bdg} compatible with the analysis of the free energy of the system: $\e^{\iu\phi_0} = 1, \e^{\iu\phi_{1}} = \iu $ and $\e^{\iu\phi_{-1}} = \iu$.
Since we are studying topological properties, we are only concerned with $\mathcal{H}_{\v{k}}$ up to an equivalence given by smooth deformations that do not close the gap.  
To simplify, we therefore perform an adiabatic transformation on the single particle Hamiltonian by continuously shrinking $\Delta_{\uparrow\downarrow}$ to $0$.
In doing so, we keep the Hamiltonian gapped but reduce the gap from $\sqrt{2}\Delta$ to $\Delta$ \footnote{Specifically, multiplying $\Delta_{\uparrow\downarrow}$ by $(1-\zeta)$, we find that the gap is given by $\Delta \sqrt{2} \sqrt{1 + \zeta(\zeta/2 - 1)}$, which is $\neq 0$ for all $\zeta \in (0,1)$.}.
Hence, this transformation should leave the topological character of the system untouched.

Having performed such a transformation, the $U$ matrices read
\begin{align}\label{eq:U_matrices}
	U^{1} \coloneqq \rho^1 \otimes \sigma^1 \otimes \tau^1, \quad 
	U^{2} \coloneqq \rho^2 \otimes \sigma^1 \otimes \tau^1, 
	\quad \text{and} \quad
	U^{3} \coloneqq \rho^3 \otimes \sigma^0 \otimes \tau^0.
\end{align}
Importantly, these matrices satisfy the same commutation and anti-commutation relations as the Pauli matrices. 
It is straightforward to check that the inverse propagator in this case is given by 
\begin{equation}
	\left( \iu k_0 \mathbf{1} - \v{m}_{\v{k}} \cdot\v{U} \right)^{-1} = \frac{1}{(\iu k_0)^2 - \v{m}_{\v{k}}^2} \left( \iu k_0 \mathbf{1} + \v{m}_{\v{k}} \cdot \v{U} \right).
\end{equation}
Multiplying out the terms and using the trace identity $\tr\left(U^{\alpha} U^{\beta} U^{\gamma}\right) = 8\iu \epsilon^{\alpha\beta\gamma}$ yields
\begin{align}
    \mathsf{k} &= \frac{\iu}{8\pi^2} \int \D^2 k \int \D k_{0} \frac{8\iu \epsilon^{\alpha\gamma\beta}}{\left( (\iu k_0)^2 - \v{m}_{\v{k}}^2 \right)^{3}} \big[ - (\iu k_0)^2 + \v{m}_{\v{k}}^2 \big] m^{\gamma}_{\v{k}} \partial_{k_1}m^{\alpha}_{\v{k}} \partial_{k_{2}} m^{\beta}_{\v{k}} \notag \\
    &= -\frac{1}{\pi^2} \int \D^2 k \epsilon^{\alpha\beta\gamma} m^{\gamma}_{\v{k}} \partial_{k_1} m^{\alpha}_{\v{k}} \partial_{k_2}m^{\beta}_{\v{k}} \times \int \D k_0 \frac{1}{(k_0^2 + \v{m}_{\v{k}}^2)^2} \notag \\
    &= -\frac{1}{2\pi} \int \D^2 k \frac{1}{\abs{\v{m}}^3} \epsilon^{\alpha\beta\gamma} m^{\gamma}_{\v{k}} \partial_{k_1} m^{\alpha}_{\v{k}} \partial_{k_2}m^{\beta}_{\v{k}}.
\end{align}

Writing the inverse Green's function as $\mathscr{G}^{-1}(k) = \iu k_0 \mathbf{1} - \v{m}_{\v{k}} \cdot\v{U}$ with the $U$ matrices as defined in Eq.~\eqref{eq:U_matrices} implies that 
\begin{equation}
    \v{m}_{\v{k}} = \begin{pmatrix} \displaystyle
        - \frac{\Delta}{\abs{\v{k}}} k_y &  \displaystyle \frac{\Delta}{\abs{\v{k}}} k_x & \displaystyle \xi_{\v{k}}   
    \end{pmatrix}^{\mathsf{T}}.
\end{equation}
Now, recall that a central assumption of BCS theory is that $\abs{\Delta} \ll \epsilon_{\mathrm{F}}$. 
Hence, away from the Fermi sea but within the BZ, the $\v{m}$ vector essentially points in the $\hat{z}$ direction, and the integral consequently vanishes. 
We can therefore approximate the integral over BZ by taking only the contributions arising from within the Fermi sea around the two symmetry points $K$ and $K'$.
Incidentally, we find ourselves in the fortunate position of actually being able to do the remaining integral due to the simple dispersion of the low-energy fermions within the Fermi surface
\begin{align}
	\mathsf{k} &= 2\times \frac{1}{2\pi} 2\pi \int_{0}^{k_{\mathrm{F}}} \D k k \frac{\Delta^2 c_{\psi}}{k} \Big( (c_{\psi}k -\mu)^2 + \Delta^2 \Big)^{-3/2} \notag \\
	&= 2\Delta^2 c_{\psi} \int_{0}^{k_{\mathrm{F}}} \D k   \Big( (c_{\psi}k -\mu)^2 + \Delta^2 \Big)^{-3/2} \notag \\
    &= 2\frac{\mu}{\sqrt{\mu^2 + \Delta^2}} \approx 2 \sign{\mu},
\end{align} 
where we have used $c_{\psi} k_{\mathrm{F}} \equiv \mu$, and $\abs{\Delta} \ll \mu$ in the last transition.

    Recall that we at some point in the derivation assumed that $\mu>0$ to restrict to only the $+$ band of the low-energy fermions. 
    If we instead assumed $\mu < 0$, the two first components of the $\v{m}$ vector are left untouched due to the appearance of the same $g_{\v{k}}$ factors, but the third would be $\xi_{\v{k}} = - c_{\psi} \abs{\v{k}} - \mu$. 
    However, doing the integral, now from $k = 0 $ to $k = - \mu/c_{\psi}$, still yields $\mathsf{k} = 2 \sign{\mu}$, so the result persists even in this case. 

\twocolumngrid

\bibliography{references}

\begin{thebibliography}{68}%
\makeatletter
\providecommand \@ifxundefined [1]{%
 \@ifx{#1\undefined}
}%
\providecommand \@ifnum [1]{%
 \ifnum #1\expandafter \@firstoftwo
 \else \expandafter \@secondoftwo
 \fi
}%
\providecommand \@ifx [1]{%
 \ifx #1\expandafter \@firstoftwo
 \else \expandafter \@secondoftwo
 \fi
}%
\providecommand \natexlab [1]{#1}%
\providecommand \enquote  [1]{``#1''}%
\providecommand \bibnamefont  [1]{#1}%
\providecommand \bibfnamefont [1]{#1}%
\providecommand \citenamefont [1]{#1}%
\providecommand \href@noop [0]{\@secondoftwo}%
\providecommand \href [0]{\begingroup \@sanitize@url \@href}%
\providecommand \@href[1]{\@@startlink{#1}\@@href}%
\providecommand \@@href[1]{\endgroup#1\@@endlink}%
\providecommand \@sanitize@url [0]{\catcode `\\12\catcode `\$12\catcode `\&12\catcode `\#12\catcode `\^12\catcode `\_12\catcode `\%12\relax}%
\providecommand \@@startlink[1]{}%
\providecommand \@@endlink[0]{}%
\providecommand \url  [0]{\begingroup\@sanitize@url \@url }%
\providecommand \@url [1]{\endgroup\@href {#1}{\urlprefix }}%
\providecommand \urlprefix  [0]{URL }%
\providecommand \Eprint [0]{\href }%
\providecommand \doibase [0]{https://doi.org/}%
\providecommand \selectlanguage [0]{\@gobble}%
\providecommand \bibinfo  [0]{\@secondoftwo}%
\providecommand \bibfield  [0]{\@secondoftwo}%
\providecommand \translation [1]{[#1]}%
\providecommand \BibitemOpen [0]{}%
\providecommand \bibitemStop [0]{}%
\providecommand \bibitemNoStop [0]{.\EOS\space}%
\providecommand \EOS [0]{\spacefactor3000\relax}%
\providecommand \BibitemShut  [1]{\csname bibitem#1\endcsname}%
\let\auto@bib@innerbib\@empty
%</preamble>
\bibitem [{\citenamefont {Nayak}\ \emph {et~al.}(2008)\citenamefont {Nayak}, \citenamefont {Simon}, \citenamefont {Stern}, \citenamefont {Freedman},\ and\ \citenamefont {Das~Sarma}}]{nayakNonAbelianAnyonsTopological2008}%
  \BibitemOpen
  \bibfield  {author} {\bibinfo {author} {\bibfnamefont {C.}~\bibnamefont {Nayak}}, \bibinfo {author} {\bibfnamefont {S.~H.}\ \bibnamefont {Simon}}, \bibinfo {author} {\bibfnamefont {A.}~\bibnamefont {Stern}}, \bibinfo {author} {\bibfnamefont {M.}~\bibnamefont {Freedman}},\ and\ \bibinfo {author} {\bibfnamefont {S.}~\bibnamefont {Das~Sarma}},\ }\bibfield  {title} {\bibinfo {title} {Non-{{Abelian}} anyons and topological quantum computation},\ }\href {https://doi.org/10.1103/RevModPhys.80.1083} {\bibfield  {journal} {\bibinfo  {journal} {Reviews of Modern Physics}\ }\textbf {\bibinfo {volume} {80}},\ \bibinfo {pages} {1083} (\bibinfo {year} {2008})}\BibitemShut {NoStop}%
\bibitem [{\citenamefont {M{\'e}nard}\ \emph {et~al.}(2019)\citenamefont {M{\'e}nard}, \citenamefont {Mesaros}, \citenamefont {Brun}, \citenamefont {Debontridder}, \citenamefont {Roditchev}, \citenamefont {Simon},\ and\ \citenamefont {Cren}}]{menardIsolatedPairsMajorana2019}%
  \BibitemOpen
  \bibfield  {author} {\bibinfo {author} {\bibfnamefont {G.~C.}\ \bibnamefont {M{\'e}nard}}, \bibinfo {author} {\bibfnamefont {A.}~\bibnamefont {Mesaros}}, \bibinfo {author} {\bibfnamefont {C.}~\bibnamefont {Brun}}, \bibinfo {author} {\bibfnamefont {F.}~\bibnamefont {Debontridder}}, \bibinfo {author} {\bibfnamefont {D.}~\bibnamefont {Roditchev}}, \bibinfo {author} {\bibfnamefont {P.}~\bibnamefont {Simon}},\ and\ \bibinfo {author} {\bibfnamefont {T.}~\bibnamefont {Cren}},\ }\bibfield  {title} {\bibinfo {title} {Isolated pairs of {{Majorana}} zero modes in a disordered superconducting lead monolayer},\ }\href {https://doi.org/10.1038/s41467-019-10397-5} {\bibfield  {journal} {\bibinfo  {journal} {Nature Communications}\ }\textbf {\bibinfo {volume} {10}},\ \bibinfo {pages} {2587} (\bibinfo {year} {2019})}\BibitemShut {NoStop}%
\bibitem [{\citenamefont {Sato}\ and\ \citenamefont {Ando}(2017)}]{satoTopologicalSuperconductorsReview2017}%
  \BibitemOpen
  \bibfield  {author} {\bibinfo {author} {\bibfnamefont {M.}~\bibnamefont {Sato}}\ and\ \bibinfo {author} {\bibfnamefont {Y.}~\bibnamefont {Ando}},\ }\bibfield  {title} {\bibinfo {title} {Topological superconductors: A review},\ }\href {https://doi.org/10.1088/1361-6633/aa6ac7} {\bibfield  {journal} {\bibinfo  {journal} {Reports on Progress in Physics}\ }\textbf {\bibinfo {volume} {80}},\ \bibinfo {pages} {076501} (\bibinfo {year} {2017})}\BibitemShut {NoStop}%
\bibitem [{\citenamefont {Zlotnikov}\ \emph {et~al.}(2021)\citenamefont {Zlotnikov}, \citenamefont {Shustin},\ and\ \citenamefont {Fedoseev}}]{zlotnikovAspectsTopologicalSuperconductivity2021}%
  \BibitemOpen
  \bibfield  {author} {\bibinfo {author} {\bibfnamefont {A.~O.}\ \bibnamefont {Zlotnikov}}, \bibinfo {author} {\bibfnamefont {M.~S.}\ \bibnamefont {Shustin}},\ and\ \bibinfo {author} {\bibfnamefont {A.~D.}\ \bibnamefont {Fedoseev}},\ }\bibfield  {title} {\bibinfo {title} {Aspects of {{Topological Superconductivity}} in {{2D Systems}}: {{Noncollinear Magnetism}}, {{Skyrmions}}, and {{Higher-order Topology}}},\ }\href {https://doi.org/10.1007/s10948-021-06029-z} {\bibfield  {journal} {\bibinfo  {journal} {Journal of Superconductivity and Novel Magnetism}\ }\textbf {\bibinfo {volume} {34}},\ \bibinfo {pages} {3053} (\bibinfo {year} {2021})}\BibitemShut {NoStop}%
\bibitem [{\citenamefont {Nakosai}\ \emph {et~al.}(2013)\citenamefont {Nakosai}, \citenamefont {Tanaka},\ and\ \citenamefont {Nagaosa}}]{nakosaiTwodimensionalWaveSuperconducting2013}%
  \BibitemOpen
  \bibfield  {author} {\bibinfo {author} {\bibfnamefont {S.}~\bibnamefont {Nakosai}}, \bibinfo {author} {\bibfnamefont {Y.}~\bibnamefont {Tanaka}},\ and\ \bibinfo {author} {\bibfnamefont {N.}~\bibnamefont {Nagaosa}},\ }\bibfield  {title} {\bibinfo {title} {Two-dimensional {$p$}-wave superconducting states with magnetic moments on a conventional {$s$}-wave superconductor},\ }\href {https://doi.org/10.1103/PhysRevB.88.180503} {\bibfield  {journal} {\bibinfo  {journal} {Physical Review B}\ }\textbf {\bibinfo {volume} {88}},\ \bibinfo {pages} {180503} (\bibinfo {year} {2013})}\BibitemShut {NoStop}%
\bibitem [{\citenamefont {Chen}\ and\ \citenamefont {Schnyder}(2015)}]{chenMajoranaEdgeStates2015}%
  \BibitemOpen
  \bibfield  {author} {\bibinfo {author} {\bibfnamefont {W.}~\bibnamefont {Chen}}\ and\ \bibinfo {author} {\bibfnamefont {A.~P.}\ \bibnamefont {Schnyder}},\ }\bibfield  {title} {\bibinfo {title} {Majorana edge states in superconductor-noncollinear magnet interfaces},\ }\href {https://doi.org/10.1103/PhysRevB.92.214502} {\bibfield  {journal} {\bibinfo  {journal} {Physical Review B}\ }\textbf {\bibinfo {volume} {92}},\ \bibinfo {pages} {214502} (\bibinfo {year} {2015})}\BibitemShut {NoStop}%
\bibitem [{\citenamefont {Rex}\ \emph {et~al.}(2019)\citenamefont {Rex}, \citenamefont {Gornyi},\ and\ \citenamefont {Mirlin}}]{rexMajoranaBoundStates2019}%
  \BibitemOpen
  \bibfield  {author} {\bibinfo {author} {\bibfnamefont {S.}~\bibnamefont {Rex}}, \bibinfo {author} {\bibfnamefont {I.~V.}\ \bibnamefont {Gornyi}},\ and\ \bibinfo {author} {\bibfnamefont {A.~D.}\ \bibnamefont {Mirlin}},\ }\bibfield  {title} {\bibinfo {title} {Majorana bound states in magnetic skyrmions imposed onto a superconductor},\ }\href {https://doi.org/10.1103/PhysRevB.100.064504} {\bibfield  {journal} {\bibinfo  {journal} {Physical Review B}\ }\textbf {\bibinfo {volume} {100}},\ \bibinfo {pages} {064504} (\bibinfo {year} {2019})}\BibitemShut {NoStop}%
\bibitem [{\citenamefont {M{\ae}land}\ and\ \citenamefont {Sudb{\o}}(2023)}]{maelandTopologicalSuperconductivityMediated2023}%
  \BibitemOpen
  \bibfield  {author} {\bibinfo {author} {\bibfnamefont {K.}~\bibnamefont {M{\ae}land}}\ and\ \bibinfo {author} {\bibfnamefont {A.}~\bibnamefont {Sudb{\o}}},\ }\bibfield  {title} {\bibinfo {title} {Topological {{Superconductivity Mediated}} by {{Skyrmionic Magnons}}},\ }\href {https://doi.org/10.1103/PhysRevLett.130.156002} {\bibfield  {journal} {\bibinfo  {journal} {Physical Review Letters}\ }\textbf {\bibinfo {volume} {130}},\ \bibinfo {pages} {156002} (\bibinfo {year} {2023})}\BibitemShut {NoStop}%
\bibitem [{\citenamefont {Anderson}(1973)}]{andersonResonatingValenceBonds1973}%
  \BibitemOpen
  \bibfield  {author} {\bibinfo {author} {\bibfnamefont {P.~W.}\ \bibnamefont {Anderson}},\ }\bibfield  {title} {\bibinfo {title} {Resonating valence bonds: {{A}} new kind of insulator?},\ }\href {https://doi.org/10.1016/0025-5408(73)90167-0} {\bibfield  {journal} {\bibinfo  {journal} {Materials Research Bulletin}\ }\textbf {\bibinfo {volume} {8}},\ \bibinfo {pages} {153} (\bibinfo {year} {1973})}\BibitemShut {NoStop}%
\bibitem [{\citenamefont {Kalmeyer}\ and\ \citenamefont {Laughlin}(1987)}]{kalmeyerEquivalenceResonatingvalencebondFractional1987}%
  \BibitemOpen
  \bibfield  {author} {\bibinfo {author} {\bibfnamefont {V.}~\bibnamefont {Kalmeyer}}\ and\ \bibinfo {author} {\bibfnamefont {R.~B.}\ \bibnamefont {Laughlin}},\ }\bibfield  {title} {\bibinfo {title} {Equivalence of the resonating-valence-bond and fractional quantum {{Hall}} states},\ }\href {https://doi.org/10.1103/PhysRevLett.59.2095} {\bibfield  {journal} {\bibinfo  {journal} {Physical Review Letters}\ }\textbf {\bibinfo {volume} {59}},\ \bibinfo {pages} {2095} (\bibinfo {year} {1987})}\BibitemShut {NoStop}%
\bibitem [{\citenamefont {Wen}\ \emph {et~al.}(1989)\citenamefont {Wen}, \citenamefont {Wilczek},\ and\ \citenamefont {Zee}}]{wenChiralSpinStates1989}%
  \BibitemOpen
  \bibfield  {author} {\bibinfo {author} {\bibfnamefont {X.~G.}\ \bibnamefont {Wen}}, \bibinfo {author} {\bibfnamefont {F.}~\bibnamefont {Wilczek}},\ and\ \bibinfo {author} {\bibfnamefont {A.}~\bibnamefont {Zee}},\ }\bibfield  {title} {\bibinfo {title} {Chiral spin states and superconductivity},\ }\href {https://doi.org/10.1103/PhysRevB.39.11413} {\bibfield  {journal} {\bibinfo  {journal} {Physical Review B}\ }\textbf {\bibinfo {volume} {39}},\ \bibinfo {pages} {11413} (\bibinfo {year} {1989})}\BibitemShut {NoStop}%
\bibitem [{\citenamefont {Kitaev}(2006)}]{kitaevAnyonsExactlySolved2006}%
  \BibitemOpen
  \bibfield  {author} {\bibinfo {author} {\bibfnamefont {A.}~\bibnamefont {Kitaev}},\ }\bibfield  {title} {\bibinfo {title} {Anyons in an exactly solved model and beyond},\ }\href {https://doi.org/10.1016/j.aop.2005.10.005} {\bibfield  {journal} {\bibinfo  {journal} {Annals of Physics}\ }\bibinfo {series} {January {{Special Issue}}},\ \textbf {\bibinfo {volume} {321}},\ \bibinfo {pages} {2} (\bibinfo {year} {2006})}\BibitemShut {NoStop}%
\bibitem [{\citenamefont {Wen}(1991)}]{wenMeanfieldTheorySpinliquid1991a}%
  \BibitemOpen
  \bibfield  {author} {\bibinfo {author} {\bibfnamefont {X.~G.}\ \bibnamefont {Wen}},\ }\bibfield  {title} {\bibinfo {title} {Mean-field theory of spin-liquid states with finite energy gap and topological orders},\ }\href {https://doi.org/10.1103/PhysRevB.44.2664} {\bibfield  {journal} {\bibinfo  {journal} {Physical Review B}\ }\textbf {\bibinfo {volume} {44}},\ \bibinfo {pages} {2664} (\bibinfo {year} {1991})}\BibitemShut {NoStop}%
\bibitem [{\citenamefont {Wen}\ and\ \citenamefont {Zee}(1989)}]{wenEffectiveTheoryPbreaking1989}%
  \BibitemOpen
  \bibfield  {author} {\bibinfo {author} {\bibfnamefont {X.~G.}\ \bibnamefont {Wen}}\ and\ \bibinfo {author} {\bibfnamefont {A.}~\bibnamefont {Zee}},\ }\bibfield  {title} {\bibinfo {title} {Effective theory of the {{T-}} and {{P-breaking}} superconducting state},\ }\href {https://doi.org/10.1103/PhysRevLett.62.2873} {\bibfield  {journal} {\bibinfo  {journal} {Physical Review Letters}\ }\textbf {\bibinfo {volume} {62}},\ \bibinfo {pages} {2873} (\bibinfo {year} {1989})}\BibitemShut {NoStop}%
\bibitem [{\citenamefont {Seifert}\ \emph {et~al.}(2018)\citenamefont {Seifert}, \citenamefont {Meng},\ and\ \citenamefont {Vojta}}]{seifertFractionalizedFermiLiquids2018b}%
  \BibitemOpen
  \bibfield  {author} {\bibinfo {author} {\bibfnamefont {U.~F.~P.}\ \bibnamefont {Seifert}}, \bibinfo {author} {\bibfnamefont {T.}~\bibnamefont {Meng}},\ and\ \bibinfo {author} {\bibfnamefont {M.}~\bibnamefont {Vojta}},\ }\bibfield  {title} {\bibinfo {title} {Fractionalized {{Fermi}} liquids and exotic superconductivity in the {{Kitaev-Kondo}} lattice},\ }\href {https://doi.org/10.1103/PhysRevB.97.085118} {\bibfield  {journal} {\bibinfo  {journal} {Physical Review B}\ }\textbf {\bibinfo {volume} {97}},\ \bibinfo {pages} {085118} (\bibinfo {year} {2018})}\BibitemShut {NoStop}%
\bibitem [{\citenamefont {Choi}\ \emph {et~al.}(2018)\citenamefont {Choi}, \citenamefont {Klein}, \citenamefont {Rosch},\ and\ \citenamefont {Kim}}]{choiTopologicalSuperconductivityKondoKitaev2018}%
  \BibitemOpen
  \bibfield  {author} {\bibinfo {author} {\bibfnamefont {W.}~\bibnamefont {Choi}}, \bibinfo {author} {\bibfnamefont {P.~W.}\ \bibnamefont {Klein}}, \bibinfo {author} {\bibfnamefont {A.}~\bibnamefont {Rosch}},\ and\ \bibinfo {author} {\bibfnamefont {Y.~B.}\ \bibnamefont {Kim}},\ }\bibfield  {title} {\bibinfo {title} {Topological superconductivity in the {{Kondo-Kitaev}} model},\ }\href {https://doi.org/10.1103/PhysRevB.98.155123} {\bibfield  {journal} {\bibinfo  {journal} {Physical Review B}\ }\textbf {\bibinfo {volume} {98}},\ \bibinfo {pages} {155123} (\bibinfo {year} {2018})}\BibitemShut {NoStop}%
\bibitem [{\citenamefont {{de Carvalho}}\ \emph {et~al.}(2021)\citenamefont {{de Carvalho}}, \citenamefont {Teixeira}, \citenamefont {Freire},\ and\ \citenamefont {Miranda}}]{decarvalhoOddfrequencyPairDensity2021}%
  \BibitemOpen
  \bibfield  {author} {\bibinfo {author} {\bibfnamefont {V.~S.}\ \bibnamefont {{de Carvalho}}}, \bibinfo {author} {\bibfnamefont {R.~M.~P.}\ \bibnamefont {Teixeira}}, \bibinfo {author} {\bibfnamefont {H.}~\bibnamefont {Freire}},\ and\ \bibinfo {author} {\bibfnamefont {E.}~\bibnamefont {Miranda}},\ }\bibfield  {title} {\bibinfo {title} {Odd-frequency pair density wave in the {{Kitaev-Kondo}} lattice model},\ }\href {https://doi.org/10.1103/PhysRevB.103.174512} {\bibfield  {journal} {\bibinfo  {journal} {Physical Review B}\ }\textbf {\bibinfo {volume} {103}},\ \bibinfo {pages} {174512} (\bibinfo {year} {2021})}\BibitemShut {NoStop}%
\bibitem [{\citenamefont {Coleman}\ \emph {et~al.}(2022)\citenamefont {Coleman}, \citenamefont {Panigrahi},\ and\ \citenamefont {Tsvelik}}]{colemanSolvable3DKondo2022}%
  \BibitemOpen
  \bibfield  {author} {\bibinfo {author} {\bibfnamefont {P.}~\bibnamefont {Coleman}}, \bibinfo {author} {\bibfnamefont {A.}~\bibnamefont {Panigrahi}},\ and\ \bibinfo {author} {\bibfnamefont {A.}~\bibnamefont {Tsvelik}},\ }\bibfield  {title} {\bibinfo {title} {Solvable {{3D Kondo Lattice Exhibiting Pair Density Wave}}, {{Odd-Frequency Pairing}}, and {{Order Fractionalization}}},\ }\href {https://doi.org/10.1103/PhysRevLett.129.177601} {\bibfield  {journal} {\bibinfo  {journal} {Physical Review Letters}\ }\textbf {\bibinfo {volume} {129}},\ \bibinfo {pages} {177601} (\bibinfo {year} {2022})}\BibitemShut {NoStop}%
\bibitem [{\citenamefont {Tsvelik}\ and\ \citenamefont {Coleman}(2022)}]{tsvelikOrderFractionalizationKitaevKondo2022}%
  \BibitemOpen
  \bibfield  {author} {\bibinfo {author} {\bibfnamefont {A.~M.}\ \bibnamefont {Tsvelik}}\ and\ \bibinfo {author} {\bibfnamefont {P.}~\bibnamefont {Coleman}},\ }\bibfield  {title} {\bibinfo {title} {Order fractionalization in a {{Kitaev-Kondo}} model},\ }\href {https://doi.org/10.1103/PhysRevB.106.125144} {\bibfield  {journal} {\bibinfo  {journal} {Physical Review B}\ }\textbf {\bibinfo {volume} {106}},\ \bibinfo {pages} {125144} (\bibinfo {year} {2022})}\BibitemShut {NoStop}%
\bibitem [{\citenamefont {Senthil}\ \emph {et~al.}(2003)\citenamefont {Senthil}, \citenamefont {Sachdev},\ and\ \citenamefont {Vojta}}]{senthilFractionalizedFermiLiquids2003}%
  \BibitemOpen
  \bibfield  {author} {\bibinfo {author} {\bibfnamefont {T.}~\bibnamefont {Senthil}}, \bibinfo {author} {\bibfnamefont {S.}~\bibnamefont {Sachdev}},\ and\ \bibinfo {author} {\bibfnamefont {M.}~\bibnamefont {Vojta}},\ }\bibfield  {title} {\bibinfo {title} {Fractionalized {{Fermi Liquids}}},\ }\href {https://doi.org/10.1103/PhysRevLett.90.216403} {\bibfield  {journal} {\bibinfo  {journal} {Physical Review Letters}\ }\textbf {\bibinfo {volume} {90}},\ \bibinfo {pages} {216403} (\bibinfo {year} {2003})}\BibitemShut {NoStop}%
\bibitem [{\citenamefont {Kargarian}\ \emph {et~al.}(2016)\citenamefont {Kargarian}, \citenamefont {Efimkin},\ and\ \citenamefont {Galitski}}]{kargarianAmpereanPairingSurface2016}%
  \BibitemOpen
  \bibfield  {author} {\bibinfo {author} {\bibfnamefont {M.}~\bibnamefont {Kargarian}}, \bibinfo {author} {\bibfnamefont {D.~K.}\ \bibnamefont {Efimkin}},\ and\ \bibinfo {author} {\bibfnamefont {V.}~\bibnamefont {Galitski}},\ }\bibfield  {title} {\bibinfo {title} {Amperean {{Pairing}} at the {{Surface}} of {{Topological Insulators}}},\ }\href {https://doi.org/10.1103/PhysRevLett.117.076806} {\bibfield  {journal} {\bibinfo  {journal} {Physical Review Letters}\ }\textbf {\bibinfo {volume} {117}},\ \bibinfo {pages} {076806} (\bibinfo {year} {2016})}\BibitemShut {NoStop}%
\bibitem [{\citenamefont {Rohling}\ \emph {et~al.}(2018)\citenamefont {Rohling}, \citenamefont {Fj\ae{}rbu},\ and\ \citenamefont {Brataas}}]{Rohling_2018}%
  \BibitemOpen
  \bibfield  {author} {\bibinfo {author} {\bibfnamefont {N.}~\bibnamefont {Rohling}}, \bibinfo {author} {\bibfnamefont {E.~L.}\ \bibnamefont {Fj\ae{}rbu}},\ and\ \bibinfo {author} {\bibfnamefont {A.}~\bibnamefont {Brataas}},\ }\bibfield  {title} {\bibinfo {title} {Superconductivity induced by interfacial coupling to magnons},\ }\href {https://doi.org/10.1103/PhysRevB.97.115401} {\bibfield  {journal} {\bibinfo  {journal} {Phys. Rev. B}\ }\textbf {\bibinfo {volume} {97}},\ \bibinfo {pages} {115401} (\bibinfo {year} {2018})}\BibitemShut {NoStop}%
\bibitem [{\citenamefont {Hugdal}\ \emph {et~al.}(2018)\citenamefont {Hugdal}, \citenamefont {Rex}, \citenamefont {Nogueira},\ and\ \citenamefont {Sudb{\o}}}]{hugdalMagnoninducedSuperconductivityTopological2018}%
  \BibitemOpen
  \bibfield  {author} {\bibinfo {author} {\bibfnamefont {H.~G.}\ \bibnamefont {Hugdal}}, \bibinfo {author} {\bibfnamefont {S.}~\bibnamefont {Rex}}, \bibinfo {author} {\bibfnamefont {F.~S.}\ \bibnamefont {Nogueira}},\ and\ \bibinfo {author} {\bibfnamefont {A.}~\bibnamefont {Sudb{\o}}},\ }\bibfield  {title} {\bibinfo {title} {Magnon-induced superconductivity in a topological insulator coupled to ferromagnetic and antiferromagnetic insulators},\ }\href {https://doi.org/10.1103/PhysRevB.97.195438} {\bibfield  {journal} {\bibinfo  {journal} {Physical Review B}\ }\textbf {\bibinfo {volume} {97}},\ \bibinfo {pages} {195438} (\bibinfo {year} {2018})}\BibitemShut {NoStop}%
\bibitem [{\citenamefont {Erlandsen}\ \emph {et~al.}(2019)\citenamefont {Erlandsen}, \citenamefont {Kamra}, \citenamefont {Brataas},\ and\ \citenamefont {Sudb{\o}}}]{erlandsenEnhancementSuperconductivityMediated2019a}%
  \BibitemOpen
  \bibfield  {author} {\bibinfo {author} {\bibfnamefont {E.}~\bibnamefont {Erlandsen}}, \bibinfo {author} {\bibfnamefont {A.}~\bibnamefont {Kamra}}, \bibinfo {author} {\bibfnamefont {A.}~\bibnamefont {Brataas}},\ and\ \bibinfo {author} {\bibfnamefont {A.}~\bibnamefont {Sudb{\o}}},\ }\bibfield  {title} {\bibinfo {title} {Enhancement of superconductivity mediated by antiferromagnetic squeezed magnons},\ }\href {https://doi.org/10.1103/PhysRevB.100.100503} {\bibfield  {journal} {\bibinfo  {journal} {Physical Review B}\ }\textbf {\bibinfo {volume} {100}},\ \bibinfo {pages} {100503} (\bibinfo {year} {2019})}\BibitemShut {NoStop}%
\bibitem [{\citenamefont {Erlandsen}\ \emph {et~al.}(2020)\citenamefont {Erlandsen}, \citenamefont {Brataas},\ and\ \citenamefont {Sudb{\o}}}]{erlandsenMagnonmediatedSuperconductivitySurface2020}%
  \BibitemOpen
  \bibfield  {author} {\bibinfo {author} {\bibfnamefont {E.}~\bibnamefont {Erlandsen}}, \bibinfo {author} {\bibfnamefont {A.}~\bibnamefont {Brataas}},\ and\ \bibinfo {author} {\bibfnamefont {A.}~\bibnamefont {Sudb{\o}}},\ }\bibfield  {title} {\bibinfo {title} {Magnon-mediated superconductivity on the surface of a topological insulator},\ }\href {https://doi.org/10.1103/PhysRevB.101.094503} {\bibfield  {journal} {\bibinfo  {journal} {Physical Review B}\ }\textbf {\bibinfo {volume} {101}},\ \bibinfo {pages} {094503} (\bibinfo {year} {2020})}\BibitemShut {NoStop}%
\bibitem [{\citenamefont {Thingstad}\ \emph {et~al.}(2021)\citenamefont {Thingstad}, \citenamefont {Erlandsen},\ and\ \citenamefont {Sudb{\o}}}]{thingstadEliashbergStudySuperconductivity2021}%
  \BibitemOpen
  \bibfield  {author} {\bibinfo {author} {\bibfnamefont {E.}~\bibnamefont {Thingstad}}, \bibinfo {author} {\bibfnamefont {E.}~\bibnamefont {Erlandsen}},\ and\ \bibinfo {author} {\bibfnamefont {A.}~\bibnamefont {Sudb{\o}}},\ }\bibfield  {title} {\bibinfo {title} {Eliashberg study of superconductivity induced by interfacial coupling to antiferromagnets},\ }\href {https://doi.org/10.1103/PhysRevB.104.014508} {\bibfield  {journal} {\bibinfo  {journal} {Physical Review B}\ }\textbf {\bibinfo {volume} {104}},\ \bibinfo {pages} {014508} (\bibinfo {year} {2021})}\BibitemShut {NoStop}%
\bibitem [{\citenamefont {Coleman}\ and\ \citenamefont {Andrei}(1989)}]{colemanKondostabilisedSpinLiquids1989}%
  \BibitemOpen
  \bibfield  {author} {\bibinfo {author} {\bibfnamefont {P.}~\bibnamefont {Coleman}}\ and\ \bibinfo {author} {\bibfnamefont {N.}~\bibnamefont {Andrei}},\ }\bibfield  {title} {\bibinfo {title} {Kondo-stabilised spin liquids and heavy fermion superconductivity},\ }\href {https://doi.org/10.1088/0953-8984/1/26/003} {\bibfield  {journal} {\bibinfo  {journal} {Journal of Physics: Condensed Matter}\ }\textbf {\bibinfo {volume} {1}},\ \bibinfo {pages} {4057} (\bibinfo {year} {1989})}\BibitemShut {NoStop}%
\bibitem [{\citenamefont {Chatterjee}\ \emph {et~al.}(2016)\citenamefont {Chatterjee}, \citenamefont {Qi}, \citenamefont {Sachdev},\ and\ \citenamefont {Steinberg}}]{chatterjeeSuperconductivityConfinementTransition2016}%
  \BibitemOpen
  \bibfield  {author} {\bibinfo {author} {\bibfnamefont {S.}~\bibnamefont {Chatterjee}}, \bibinfo {author} {\bibfnamefont {Y.}~\bibnamefont {Qi}}, \bibinfo {author} {\bibfnamefont {S.}~\bibnamefont {Sachdev}},\ and\ \bibinfo {author} {\bibfnamefont {J.}~\bibnamefont {Steinberg}},\ }\bibfield  {title} {\bibinfo {title} {Superconductivity from a confinement transition out of a fractionalized fermi liquid with {$\mathbb{Z}_2$} topological and ising-nematic orders},\ }\href {https://doi.org/10.1103/PhysRevB.94.024502} {\bibfield  {journal} {\bibinfo  {journal} {Physical Review B}\ }\textbf {\bibinfo {volume} {94}},\ \bibinfo {pages} {024502} (\bibinfo {year} {2016})}\BibitemShut {NoStop}%
\bibitem [{\citenamefont {Imry}(1975)}]{imryStatisticalMechanicsCoupled1975a}%
  \BibitemOpen
  \bibfield  {author} {\bibinfo {author} {\bibfnamefont {Y.}~\bibnamefont {Imry}},\ }\bibfield  {title} {\bibinfo {title} {On the statistical mechanics of coupled order parameters},\ }\href {https://doi.org/10.1088/0022-3719/8/5/005} {\bibfield  {journal} {\bibinfo  {journal} {Journal of Physics C: Solid State Physics}\ }\textbf {\bibinfo {volume} {8}},\ \bibinfo {pages} {567} (\bibinfo {year} {1975})}\BibitemShut {NoStop}%
\bibitem [{\citenamefont {Bruce}\ and\ \citenamefont {Aharony}(1975)}]{bruceCoupledOrderParameters1975a}%
  \BibitemOpen
  \bibfield  {author} {\bibinfo {author} {\bibfnamefont {A.~D.}\ \bibnamefont {Bruce}}\ and\ \bibinfo {author} {\bibfnamefont {A.}~\bibnamefont {Aharony}},\ }\bibfield  {title} {\bibinfo {title} {Coupled order parameters, symmetry-breaking irrelevant scaling fields, and tetracritical points},\ }\href {https://doi.org/10.1103/PhysRevB.11.478} {\bibfield  {journal} {\bibinfo  {journal} {Physical Review B}\ }\textbf {\bibinfo {volume} {11}},\ \bibinfo {pages} {478} (\bibinfo {year} {1975})}\BibitemShut {NoStop}%
\bibitem [{\citenamefont {Calabrese}\ \emph {et~al.}(2003)\citenamefont {Calabrese}, \citenamefont {Pelissetto},\ and\ \citenamefont {Vicari}}]{calabreseMulticriticalPhenomenaMathrm2003a}%
  \BibitemOpen
  \bibfield  {author} {\bibinfo {author} {\bibfnamefont {P.}~\bibnamefont {Calabrese}}, \bibinfo {author} {\bibfnamefont {A.}~\bibnamefont {Pelissetto}},\ and\ \bibinfo {author} {\bibfnamefont {E.}~\bibnamefont {Vicari}},\ }\bibfield  {title} {\bibinfo {title} {Multicritical phenomena in $\mathrm{O}{(n}_{1})\ensuremath{\bigoplus}\mathrm{O}{(n}_{2})$-symmetric theories},\ }\href {https://doi.org/10.1103/PhysRevB.67.054505} {\bibfield  {journal} {\bibinfo  {journal} {Physical Review B}\ }\textbf {\bibinfo {volume} {67}},\ \bibinfo {pages} {054505} (\bibinfo {year} {2003})}\BibitemShut {NoStop}%
\bibitem [{\citenamefont {You}\ \emph {et~al.}(2012)\citenamefont {You}, \citenamefont {Kimchi},\ and\ \citenamefont {Vishwanath}}]{youDopingSpinorbitMott2012}%
  \BibitemOpen
  \bibfield  {author} {\bibinfo {author} {\bibfnamefont {Y.-Z.}\ \bibnamefont {You}}, \bibinfo {author} {\bibfnamefont {I.}~\bibnamefont {Kimchi}},\ and\ \bibinfo {author} {\bibfnamefont {A.}~\bibnamefont {Vishwanath}},\ }\bibfield  {title} {\bibinfo {title} {Doping a spin-orbit mott insulator: Topological superconductivity from the kitaev-heisenberg model and possible application to (na${}_{2}$/li${}_{2}$)iro${}_{3}$},\ }\href {https://doi.org/10.1103/PhysRevB.86.085145} {\bibfield  {journal} {\bibinfo  {journal} {Physical Review B}\ }\textbf {\bibinfo {volume} {86}},\ \bibinfo {pages} {085145} (\bibinfo {year} {2012})}\BibitemShut {NoStop}%
\bibitem [{\citenamefont {Abrikosov}(1965)}]{Abrikosov1965}%
  \BibitemOpen
  \bibfield  {author} {\bibinfo {author} {\bibfnamefont {A.~A.}\ \bibnamefont {Abrikosov}},\ }\bibfield  {title} {\bibinfo {title} {Electron scattering on magnetic impurities in metals and anomalous resistivity effects},\ }\href {https://doi.org/10.1103/PhysicsPhysiqueFizika.2.5} {\bibfield  {journal} {\bibinfo  {journal} {Physics Physique Fizika}\ }\textbf {\bibinfo {volume} {2}},\ \bibinfo {pages} {5} (\bibinfo {year} {1965})}\BibitemShut {NoStop}%
\bibitem [{\citenamefont {Affleck}\ \emph {et~al.}(1988)\citenamefont {Affleck}, \citenamefont {Zou}, \citenamefont {Hsu},\ and\ \citenamefont {Anderson}}]{affleckSUGaugeSymmetry1988}%
  \BibitemOpen
  \bibfield  {author} {\bibinfo {author} {\bibfnamefont {I.}~\bibnamefont {Affleck}}, \bibinfo {author} {\bibfnamefont {Z.}~\bibnamefont {Zou}}, \bibinfo {author} {\bibfnamefont {T.}~\bibnamefont {Hsu}},\ and\ \bibinfo {author} {\bibfnamefont {P.~W.}\ \bibnamefont {Anderson}},\ }\bibfield  {title} {\bibinfo {title} {{{SU}}(2) gauge symmetry of the large-${{U}}$ limit of the {{Hubbard}} model},\ }\href {https://doi.org/10.1103/PhysRevB.38.745} {\bibfield  {journal} {\bibinfo  {journal} {Physical Review B}\ }\textbf {\bibinfo {volume} {38}},\ \bibinfo {pages} {745} (\bibinfo {year} {1988})}\BibitemShut {NoStop}%
\bibitem [{\citenamefont {Tsvelik}(1992)}]{tsvelikNewFermionicDescription1992}%
  \BibitemOpen
  \bibfield  {author} {\bibinfo {author} {\bibfnamefont {A.~M.}\ \bibnamefont {Tsvelik}},\ }\bibfield  {title} {\bibinfo {title} {New fermionic description of quantum spin liquid state},\ }\href {https://doi.org/10.1103/PhysRevLett.69.2142} {\bibfield  {journal} {\bibinfo  {journal} {Physical Review Letters}\ }\textbf {\bibinfo {volume} {69}},\ \bibinfo {pages} {2142} (\bibinfo {year} {1992})}\BibitemShut {NoStop}%
\bibitem [{Note1()}]{Note1}%
  \BibitemOpen
  \bibinfo {note} {As argued in previous studies, these turn out to vanish at the mean-field level anyway \cite {youDopingSpinorbitMott2012,seifertFractionalizedFermiLiquids2018b}}\BibitemShut {NoStop}%
\bibitem [{\citenamefont {Kogan}(2011)}]{koganRKKYInteractionGraphene2011}%
  \BibitemOpen
  \bibfield  {author} {\bibinfo {author} {\bibfnamefont {E.}~\bibnamefont {Kogan}},\ }\bibfield  {title} {\bibinfo {title} {{{RKKY}} interaction in graphene},\ }\href {https://doi.org/10.1103/PhysRevB.84.115119} {\bibfield  {journal} {\bibinfo  {journal} {Physical Review B}\ }\textbf {\bibinfo {volume} {84}},\ \bibinfo {pages} {115119} (\bibinfo {year} {2011})}\BibitemShut {NoStop}%
\bibitem [{Note2()}]{Note2}%
  \BibitemOpen
  \bibinfo {note} {We comment on the case of $s=-$ at a later stage.}\BibitemShut {Stop}%
\bibitem [{\citenamefont {Hohenberg}(1967)}]{Hohenberg}%
  \BibitemOpen
  \bibfield  {author} {\bibinfo {author} {\bibfnamefont {P.~C.}\ \bibnamefont {Hohenberg}},\ }\bibfield  {title} {\bibinfo {title} {Existence of long-range order in one and two dimensions},\ }\href {https://doi.org/10.1103/PhysRev.158.383} {\bibfield  {journal} {\bibinfo  {journal} {Phys. Rev.}\ }\textbf {\bibinfo {volume} {158}},\ \bibinfo {pages} {383} (\bibinfo {year} {1967})}\BibitemShut {NoStop}%
\bibitem [{\citenamefont {Mermin}\ and\ \citenamefont {Wagner}(1966)}]{MerminWagner}%
  \BibitemOpen
  \bibfield  {author} {\bibinfo {author} {\bibfnamefont {N.~D.}\ \bibnamefont {Mermin}}\ and\ \bibinfo {author} {\bibfnamefont {H.}~\bibnamefont {Wagner}},\ }\bibfield  {title} {\bibinfo {title} {Absence of ferromagnetism or antiferromagnetism in one- or two-dimensional isotropic heisenberg models},\ }\href {https://doi.org/10.1103/PhysRevLett.17.1133} {\bibfield  {journal} {\bibinfo  {journal} {Phys. Rev. Lett.}\ }\textbf {\bibinfo {volume} {17}},\ \bibinfo {pages} {1133} (\bibinfo {year} {1966})}\BibitemShut {NoStop}%
\bibitem [{\citenamefont {Kosterlitz}\ and\ \citenamefont {Thouless}(1973)}]{kosterlitzOrderingMetastabilityPhase1973}%
  \BibitemOpen
  \bibfield  {author} {\bibinfo {author} {\bibfnamefont {J.~M.}\ \bibnamefont {Kosterlitz}}\ and\ \bibinfo {author} {\bibfnamefont {D.~J.}\ \bibnamefont {Thouless}},\ }\bibfield  {title} {\bibinfo {title} {Ordering, metastability and phase transitions in two-dimensional systems},\ }\href {https://doi.org/10.1088/0022-3719/6/7/010} {\bibfield  {journal} {\bibinfo  {journal} {Journal of Physics C: Solid State Physics}\ }\textbf {\bibinfo {volume} {6}},\ \bibinfo {pages} {1181} (\bibinfo {year} {1973})}\BibitemShut {NoStop}%
\bibitem [{\citenamefont {Nelson}\ and\ \citenamefont {Kosterlitz}(1977)}]{NelsonKosterlitz}%
  \BibitemOpen
  \bibfield  {author} {\bibinfo {author} {\bibfnamefont {D.~R.}\ \bibnamefont {Nelson}}\ and\ \bibinfo {author} {\bibfnamefont {J.~M.}\ \bibnamefont {Kosterlitz}},\ }\bibfield  {title} {\bibinfo {title} {Universal jump in the superfluid density of two-dimensional superfluids},\ }\href {https://doi.org/10.1103/PhysRevLett.39.1201} {\bibfield  {journal} {\bibinfo  {journal} {Phys. Rev. Lett.}\ }\textbf {\bibinfo {volume} {39}},\ \bibinfo {pages} {1201} (\bibinfo {year} {1977})}\BibitemShut {NoStop}%
\bibitem [{\citenamefont {Loktev}\ and\ \citenamefont {Turkowski}(2009)}]{loktevSuppressionSuperconductingTransition2009}%
  \BibitemOpen
  \bibfield  {author} {\bibinfo {author} {\bibfnamefont {V.~M.}\ \bibnamefont {Loktev}}\ and\ \bibinfo {author} {\bibfnamefont {V.}~\bibnamefont {Turkowski}},\ }\bibfield  {title} {\bibinfo {title} {Suppression of the superconducting transition temperature of doped graphene due to thermal fluctuations of the order parameter},\ }\href {https://doi.org/10.1103/PhysRevB.79.233402} {\bibfield  {journal} {\bibinfo  {journal} {Physical Review B}\ }\textbf {\bibinfo {volume} {79}},\ \bibinfo {pages} {233402} (\bibinfo {year} {2009})}\BibitemShut {NoStop}%
\bibitem [{\citenamefont {Ng}\ and\ \citenamefont {Nagaosa}(2009)}]{ngBrokenTimereversalSymmetry2009}%
  \BibitemOpen
  \bibfield  {author} {\bibinfo {author} {\bibfnamefont {T.~K.}\ \bibnamefont {Ng}}\ and\ \bibinfo {author} {\bibfnamefont {N.}~\bibnamefont {Nagaosa}},\ }\bibfield  {title} {\bibinfo {title} {Broken time-reversal symmetry in {{Josephson}} junction involving two-band superconductors},\ }\href {https://doi.org/10.1209/0295-5075/87/17003} {\bibfield  {journal} {\bibinfo  {journal} {Europhysics Letters}\ }\textbf {\bibinfo {volume} {87}},\ \bibinfo {pages} {17003} (\bibinfo {year} {2009})}\BibitemShut {NoStop}%
\bibitem [{\citenamefont {Bojesen}\ \emph {et~al.}(2013)\citenamefont {Bojesen}, \citenamefont {Babaev},\ and\ \citenamefont {Sudb{\o}}}]{bojesenTimeReversalSymmetry2013}%
  \BibitemOpen
  \bibfield  {author} {\bibinfo {author} {\bibfnamefont {T.~A.}\ \bibnamefont {Bojesen}}, \bibinfo {author} {\bibfnamefont {E.}~\bibnamefont {Babaev}},\ and\ \bibinfo {author} {\bibfnamefont {A.}~\bibnamefont {Sudb{\o}}},\ }\bibfield  {title} {\bibinfo {title} {Time reversal symmetry breakdown in normal and superconducting states in frustrated three-band systems},\ }\href {https://doi.org/10.1103/PhysRevB.88.220511} {\bibfield  {journal} {\bibinfo  {journal} {Physical Review B}\ }\textbf {\bibinfo {volume} {88}},\ \bibinfo {pages} {220511} (\bibinfo {year} {2013})}\BibitemShut {NoStop}%
\bibitem [{\citenamefont {Bojesen}\ \emph {et~al.}(2014)\citenamefont {Bojesen}, \citenamefont {Babaev},\ and\ \citenamefont {Sudb{\o}}}]{bojesenPhaseTransitionsAnomalous2014}%
  \BibitemOpen
  \bibfield  {author} {\bibinfo {author} {\bibfnamefont {T.~A.}\ \bibnamefont {Bojesen}}, \bibinfo {author} {\bibfnamefont {E.}~\bibnamefont {Babaev}},\ and\ \bibinfo {author} {\bibfnamefont {A.}~\bibnamefont {Sudb{\o}}},\ }\bibfield  {title} {\bibinfo {title} {Phase transitions and anomalous normal state in superconductors with broken time-reversal symmetry},\ }\href {https://doi.org/10.1103/PhysRevB.89.104509} {\bibfield  {journal} {\bibinfo  {journal} {Physical Review B}\ }\textbf {\bibinfo {volume} {89}},\ \bibinfo {pages} {104509} (\bibinfo {year} {2014})}\BibitemShut {NoStop}%
\bibitem [{\citenamefont {Bardeen}\ \emph {et~al.}(1957)\citenamefont {Bardeen}, \citenamefont {Cooper},\ and\ \citenamefont {Schrieffer}}]{bardeenTheorySuperconductivity1957a}%
  \BibitemOpen
  \bibfield  {author} {\bibinfo {author} {\bibfnamefont {J.}~\bibnamefont {Bardeen}}, \bibinfo {author} {\bibfnamefont {L.~N.}\ \bibnamefont {Cooper}},\ and\ \bibinfo {author} {\bibfnamefont {J.~R.}\ \bibnamefont {Schrieffer}},\ }\bibfield  {title} {\bibinfo {title} {Theory of {{Superconductivity}}},\ }\href {https://doi.org/10.1103/PhysRev.108.1175} {\bibfield  {journal} {\bibinfo  {journal} {Physical Review}\ }\textbf {\bibinfo {volume} {108}},\ \bibinfo {pages} {1175} (\bibinfo {year} {1957})}\BibitemShut {NoStop}%
\bibitem [{\citenamefont {Kopnin}\ and\ \citenamefont {Sonin}(2008)}]{kopninBCSSuperconductivityDirac2008}%
  \BibitemOpen
  \bibfield  {author} {\bibinfo {author} {\bibfnamefont {N.~B.}\ \bibnamefont {Kopnin}}\ and\ \bibinfo {author} {\bibfnamefont {E.~B.}\ \bibnamefont {Sonin}},\ }\bibfield  {title} {\bibinfo {title} {{{BCS Superconductivity}} of {{Dirac Electrons}} in {{Graphene Layers}}},\ }\href {https://doi.org/10.1103/PhysRevLett.100.246808} {\bibfield  {journal} {\bibinfo  {journal} {Physical Review Letters}\ }\textbf {\bibinfo {volume} {100}},\ \bibinfo {pages} {246808} (\bibinfo {year} {2008})}\BibitemShut {NoStop}%
\bibitem [{\citenamefont {Xu}\ and\ \citenamefont {Yang}(2017)}]{xuDeterminationGapSolution2017}%
  \BibitemOpen
  \bibfield  {author} {\bibinfo {author} {\bibfnamefont {C.}~\bibnamefont {Xu}}\ and\ \bibinfo {author} {\bibfnamefont {Y.}~\bibnamefont {Yang}},\ }\bibfield  {title} {\bibinfo {title} {Determination of gap solution and critical temperature in doped graphene superconductivity},\ }\href {https://doi.org/10.1007/s00033-017-0779-7} {\bibfield  {journal} {\bibinfo  {journal} {Zeitschrift f{\"u}r angewandte Mathematik und Physik}\ }\textbf {\bibinfo {volume} {68}},\ \bibinfo {pages} {34} (\bibinfo {year} {2017})}\BibitemShut {NoStop}%
\bibitem [{\citenamefont {Altland}\ and\ \citenamefont {Zirnbauer}(1997)}]{altlandNonstandardSymmetryClasses1997}%
  \BibitemOpen
  \bibfield  {author} {\bibinfo {author} {\bibfnamefont {A.}~\bibnamefont {Altland}}\ and\ \bibinfo {author} {\bibfnamefont {M.~R.}\ \bibnamefont {Zirnbauer}},\ }\bibfield  {title} {\bibinfo {title} {Nonstandard symmetry classes in mesoscopic normal-superconducting hybrid structures},\ }\href {https://doi.org/10.1103/PhysRevB.55.1142} {\bibfield  {journal} {\bibinfo  {journal} {Physical Review B}\ }\textbf {\bibinfo {volume} {55}},\ \bibinfo {pages} {1142} (\bibinfo {year} {1997})}\BibitemShut {NoStop}%
\bibitem [{\citenamefont {Chiu}\ \emph {et~al.}(2016)\citenamefont {Chiu}, \citenamefont {Teo}, \citenamefont {Schnyder},\ and\ \citenamefont {Ryu}}]{chiuClassificationTopologicalQuantum2016a}%
  \BibitemOpen
  \bibfield  {author} {\bibinfo {author} {\bibfnamefont {C.-K.}\ \bibnamefont {Chiu}}, \bibinfo {author} {\bibfnamefont {J.~C.~Y.}\ \bibnamefont {Teo}}, \bibinfo {author} {\bibfnamefont {A.~P.}\ \bibnamefont {Schnyder}},\ and\ \bibinfo {author} {\bibfnamefont {S.}~\bibnamefont {Ryu}},\ }\bibfield  {title} {\bibinfo {title} {Classification of topological quantum matter with symmetries},\ }\href {https://doi.org/10.1103/RevModPhys.88.035005} {\bibfield  {journal} {\bibinfo  {journal} {Reviews of Modern Physics}\ }\textbf {\bibinfo {volume} {88}},\ \bibinfo {pages} {035005} (\bibinfo {year} {2016})}\BibitemShut {NoStop}%
\bibitem [{\citenamefont {Zhang}(1992)}]{zhangChernSimonsLandau1992}%
  \BibitemOpen
  \bibfield  {author} {\bibinfo {author} {\bibfnamefont {S.~C.}\ \bibnamefont {Zhang}},\ }\bibfield  {title} {\bibinfo {title} {The chern{\textendash}simons{\textendash}landau{\textendash}ginzburg theory of the fractional quantum hall effect},\ }\href {https://doi.org/10.1142/S0217979292000037} {\bibfield  {journal} {\bibinfo  {journal} {International Journal of Modern Physics B}\ }\textbf {\bibinfo {volume} {06}},\ \bibinfo {pages} {25} (\bibinfo {year} {1992})}\BibitemShut {NoStop}%
\bibitem [{\citenamefont {Qi}\ \emph {et~al.}(2008)\citenamefont {Qi}, \citenamefont {Hughes},\ and\ \citenamefont {Zhang}}]{qiTopologicalFieldTheory2008}%
  \BibitemOpen
  \bibfield  {author} {\bibinfo {author} {\bibfnamefont {X.-L.}\ \bibnamefont {Qi}}, \bibinfo {author} {\bibfnamefont {T.~L.}\ \bibnamefont {Hughes}},\ and\ \bibinfo {author} {\bibfnamefont {S.-C.}\ \bibnamefont {Zhang}},\ }\bibfield  {title} {\bibinfo {title} {Topological field theory of time-reversal invariant insulators},\ }\href {https://doi.org/10.1103/PhysRevB.78.195424} {\bibfield  {journal} {\bibinfo  {journal} {Physical Review B}\ }\textbf {\bibinfo {volume} {78}},\ \bibinfo {pages} {195424} (\bibinfo {year} {2008})}\BibitemShut {NoStop}%
\bibitem [{\citenamefont {Volovik}\ and\ \citenamefont {Yakovenko}(1989)}]{volovikFractionalChargeSpin1989}%
  \BibitemOpen
  \bibfield  {author} {\bibinfo {author} {\bibfnamefont {G.~E.}\ \bibnamefont {Volovik}}\ and\ \bibinfo {author} {\bibfnamefont {V.~M.}\ \bibnamefont {Yakovenko}},\ }\bibfield  {title} {\bibinfo {title} {Fractional charge, spin and statistics of solitons in superfluid {{3He}} film},\ }\href {https://doi.org/10.1088/0953-8984/1/31/025} {\bibfield  {journal} {\bibinfo  {journal} {Journal of Physics: Condensed Matter}\ }\textbf {\bibinfo {volume} {1}},\ \bibinfo {pages} {5263} (\bibinfo {year} {1989})}\BibitemShut {NoStop}%
\bibitem [{\citenamefont {Yakovenko}(1990)}]{yakovenkoChernSimonsTermsField1990}%
  \BibitemOpen
  \bibfield  {author} {\bibinfo {author} {\bibfnamefont {V.~M.}\ \bibnamefont {Yakovenko}},\ }\bibfield  {title} {\bibinfo {title} {Chern-simons terms and $n$ field in haldane's model for the quantum hall effect without landau levels},\ }\href {https://doi.org/10.1103/PhysRevLett.65.251} {\bibfield  {journal} {\bibinfo  {journal} {Physical Review Letters}\ }\textbf {\bibinfo {volume} {65}},\ \bibinfo {pages} {251} (\bibinfo {year} {1990})}\BibitemShut {NoStop}%
\bibitem [{Note3()}]{Note3}%
  \BibitemOpen
  \bibinfo {note} {That is, $\sfPsi $ and $\sfPsi ^{\dagger }$ are in fact only one independent field, made manifest through $ \sfPsi (-k)^{\protect \mathsf {T}} \rho ^{1} \otimes \protect \mathbf {1}_{4} = \sfPsi ^{\dagger }(k) $}\BibitemShut {NoStop}%
\bibitem [{\citenamefont {Volovik}(2009)}]{volovikUniverseHeliumDroplet2009}%
  \BibitemOpen
  \bibfield  {author} {\bibinfo {author} {\bibfnamefont {G.~E.}\ \bibnamefont {Volovik}},\ }\href@noop {} {\emph {\bibinfo {title} {The {{Universe}} in a {{Helium Droplet}}}}},\ International {{Series}} of {{Monographs}} on {{Physics}}\ (\bibinfo  {publisher} {{Oxford University Press}},\ \bibinfo {address} {{Oxford, New York}},\ \bibinfo {year} {2009})\BibitemShut {NoStop}%
\bibitem [{\citenamefont {Sato}\ \emph {et~al.}(2014)\citenamefont {Sato}, \citenamefont {Yamakage},\ and\ \citenamefont {Mizushima}}]{satoMirrorMajoranaZero2014}%
  \BibitemOpen
  \bibfield  {author} {\bibinfo {author} {\bibfnamefont {M.}~\bibnamefont {Sato}}, \bibinfo {author} {\bibfnamefont {A.}~\bibnamefont {Yamakage}},\ and\ \bibinfo {author} {\bibfnamefont {T.}~\bibnamefont {Mizushima}},\ }\bibfield  {title} {\bibinfo {title} {Mirror {{Majorana}} zero modes in spinful superconductors/superfluids {{Non-Abelian}} anyons in integer quantum vortices},\ }\href {https://doi.org/10.1016/j.physe.2013.07.011} {\bibfield  {journal} {\bibinfo  {journal} {Physica E: Low-dimensional Systems and Nanostructures}\ }\bibinfo {series} {Topological {{Objects}}},\ \textbf {\bibinfo {volume} {55}},\ \bibinfo {pages} {20} (\bibinfo {year} {2014})}\BibitemShut {NoStop}%
\bibitem [{\citenamefont {Volovik}(1992)}]{volovikQuantumHallChiral1992}%
  \BibitemOpen
  \bibfield  {author} {\bibinfo {author} {\bibfnamefont {G.}~\bibnamefont {Volovik}},\ }\bibfield  {title} {\bibinfo {title} {Quantum {{Hall}} and chiral edge states in thin {{3He-A}} film},\ }\href {http://jetpletters.ru/ps/1273/article_19263.shtml} {\bibfield  {journal} {\bibinfo  {journal} {Pis'ma v Zhurnal Ehksperimental'noj i Teoreticheskoj Fiziki}\ }\textbf {\bibinfo {volume} {55}},\ \bibinfo {pages} {363} (\bibinfo {year} {1992})}\BibitemShut {NoStop}%
\bibitem [{\citenamefont {Ivanov}(2001)}]{IvanovNonAbelian2001}%
  \BibitemOpen
  \bibfield  {author} {\bibinfo {author} {\bibfnamefont {D.~A.}\ \bibnamefont {Ivanov}},\ }\bibfield  {title} {\bibinfo {title} {Non-abelian statistics of half-quantum vortices in $\mathit{p}$-wave superconductors},\ }\href {https://doi.org/10.1103/PhysRevLett.86.268} {\bibfield  {journal} {\bibinfo  {journal} {Phys. Rev. Lett.}\ }\textbf {\bibinfo {volume} {86}},\ \bibinfo {pages} {268} (\bibinfo {year} {2001})}\BibitemShut {NoStop}%
\bibitem [{\citenamefont {Kawakami}\ \emph {et~al.}(2011)\citenamefont {Kawakami}, \citenamefont {Mizushima},\ and\ \citenamefont {Machida}}]{KawakamiZeroEnergyModes2011}%
  \BibitemOpen
  \bibfield  {author} {\bibinfo {author} {\bibfnamefont {T.}~\bibnamefont {Kawakami}}, \bibinfo {author} {\bibfnamefont {T.}~\bibnamefont {Mizushima}},\ and\ \bibinfo {author} {\bibfnamefont {K.}~\bibnamefont {Machida}},\ }\bibfield  {title} {\bibinfo {title} {Zero energy modes and statistics of vortices in spinful chiral p-wave superfluids},\ }\href {https://doi.org/10.1143/JPSJ.80.044603} {\bibfield  {journal} {\bibinfo  {journal} {Journal of the Physical Society of Japan}\ }\textbf {\bibinfo {volume} {80}},\ \bibinfo {pages} {044603} (\bibinfo {year} {2011})},\ \Eprint {https://arxiv.org/abs/https://doi.org/10.1143/JPSJ.80.044603} {https://doi.org/10.1143/JPSJ.80.044603} \BibitemShut {NoStop}%
\bibitem [{\citenamefont {Huang}\ \emph {et~al.}(2022)\citenamefont {Huang}, \citenamefont {Yang}, \citenamefont {Zhang},\ and\ \citenamefont {Xu}}]{huangChiralMajoranaEdge2022}%
  \BibitemOpen
  \bibfield  {author} {\bibinfo {author} {\bibfnamefont {B.}~\bibnamefont {Huang}}, \bibinfo {author} {\bibfnamefont {X.}~\bibnamefont {Yang}}, \bibinfo {author} {\bibfnamefont {Q.}~\bibnamefont {Zhang}},\ and\ \bibinfo {author} {\bibfnamefont {N.}~\bibnamefont {Xu}},\ }\bibfield  {title} {\bibinfo {title} {Chiral {{Majorana}} edge modes and vortex {{Majorana}} zero modes in superconducting antiferromagnetic topological insulator},\ }\href {https://doi.org/10.1088/1361-648X/ac4531} {\bibfield  {journal} {\bibinfo  {journal} {Journal of Physics: Condensed Matter}\ }\textbf {\bibinfo {volume} {34}},\ \bibinfo {pages} {115503} (\bibinfo {year} {2022})}\BibitemShut {NoStop}%
\bibitem [{\citenamefont {Lian}\ \emph {et~al.}(2018)\citenamefont {Lian}, \citenamefont {Sun}, \citenamefont {Vaezi}, \citenamefont {Qi},\ and\ \citenamefont {Zhang}}]{lianTopologicalQuantumComputation2018}%
  \BibitemOpen
  \bibfield  {author} {\bibinfo {author} {\bibfnamefont {B.}~\bibnamefont {Lian}}, \bibinfo {author} {\bibfnamefont {X.-Q.}\ \bibnamefont {Sun}}, \bibinfo {author} {\bibfnamefont {A.}~\bibnamefont {Vaezi}}, \bibinfo {author} {\bibfnamefont {X.-L.}\ \bibnamefont {Qi}},\ and\ \bibinfo {author} {\bibfnamefont {S.-C.}\ \bibnamefont {Zhang}},\ }\bibfield  {title} {\bibinfo {title} {Topological quantum computation based on chiral {{Majorana}} fermions},\ }\href {https://doi.org/10.1073/pnas.1810003115} {\bibfield  {journal} {\bibinfo  {journal} {Proceedings of the National Academy of Sciences}\ }\textbf {\bibinfo {volume} {115}},\ \bibinfo {pages} {10938} (\bibinfo {year} {2018})}\BibitemShut {NoStop}%
\bibitem [{\citenamefont {He}\ \emph {et~al.}(2019)\citenamefont {He}, \citenamefont {Liang}, \citenamefont {Tanaka},\ and\ \citenamefont {Nagaosa}}]{hePlatformChiralMajorana2019}%
  \BibitemOpen
  \bibfield  {author} {\bibinfo {author} {\bibfnamefont {J.~J.}\ \bibnamefont {He}}, \bibinfo {author} {\bibfnamefont {T.}~\bibnamefont {Liang}}, \bibinfo {author} {\bibfnamefont {Y.}~\bibnamefont {Tanaka}},\ and\ \bibinfo {author} {\bibfnamefont {N.}~\bibnamefont {Nagaosa}},\ }\bibfield  {title} {\bibinfo {title} {Platform of chiral {{Majorana}} edge modes and its quantum transport phenomena},\ }\href {https://doi.org/10.1038/s42005-019-0250-5} {\bibfield  {journal} {\bibinfo  {journal} {Communications Physics}\ }\textbf {\bibinfo {volume} {2}},\ \bibinfo {pages} {1} (\bibinfo {year} {2019})}\BibitemShut {NoStop}%
\bibitem [{\citenamefont {Uchoa}\ and\ \citenamefont {Castro~Neto}(2007)}]{uchoaSuperconductingStatesPure2007}%
  \BibitemOpen
  \bibfield  {author} {\bibinfo {author} {\bibfnamefont {B.}~\bibnamefont {Uchoa}}\ and\ \bibinfo {author} {\bibfnamefont {A.~H.}\ \bibnamefont {Castro~Neto}},\ }\bibfield  {title} {\bibinfo {title} {Superconducting {{States}} of {{Pure}} and {{Doped Graphene}}},\ }\href {https://doi.org/10.1103/PhysRevLett.98.146801} {\bibfield  {journal} {\bibinfo  {journal} {Physical Review Letters}\ }\textbf {\bibinfo {volume} {98}},\ \bibinfo {pages} {146801} (\bibinfo {year} {2007})}\BibitemShut {NoStop}%
\bibitem [{\citenamefont {Thingstad}\ \emph {et~al.}(2020)\citenamefont {Thingstad}, \citenamefont {Kamra}, \citenamefont {Wells},\ and\ \citenamefont {Sudb{\o}}}]{thingstadPhononmediatedSuperconductivityDoped2020}%
  \BibitemOpen
  \bibfield  {author} {\bibinfo {author} {\bibfnamefont {E.}~\bibnamefont {Thingstad}}, \bibinfo {author} {\bibfnamefont {A.}~\bibnamefont {Kamra}}, \bibinfo {author} {\bibfnamefont {J.~W.}\ \bibnamefont {Wells}},\ and\ \bibinfo {author} {\bibfnamefont {A.}~\bibnamefont {Sudb{\o}}},\ }\bibfield  {title} {\bibinfo {title} {Phonon-mediated superconductivity in doped monolayer materials},\ }\href {https://doi.org/10.1103/PhysRevB.101.214513} {\bibfield  {journal} {\bibinfo  {journal} {Physical Review B}\ }\textbf {\bibinfo {volume} {101}},\ \bibinfo {pages} {214513} (\bibinfo {year} {2020})}\BibitemShut {NoStop}%
\bibitem [{\citenamefont {Coleman}\ \emph {et~al.}(1994)\citenamefont {Coleman}, \citenamefont {Miranda},\ and\ \citenamefont {Tsvelik}}]{colemanOddfrequencyPairingKondo1994}%
  \BibitemOpen
  \bibfield  {author} {\bibinfo {author} {\bibfnamefont {P.}~\bibnamefont {Coleman}}, \bibinfo {author} {\bibfnamefont {E.}~\bibnamefont {Miranda}},\ and\ \bibinfo {author} {\bibfnamefont {A.}~\bibnamefont {Tsvelik}},\ }\bibfield  {title} {\bibinfo {title} {Odd-frequency pairing in the {{Kondo}} lattice},\ }\href {https://doi.org/10.1103/PhysRevB.49.8955} {\bibfield  {journal} {\bibinfo  {journal} {Physical Review B}\ }\textbf {\bibinfo {volume} {49}},\ \bibinfo {pages} {8955} (\bibinfo {year} {1994})}\BibitemShut {NoStop}%
\bibitem [{Note4()}]{Note4}%
  \BibitemOpen
  \bibinfo {note} {Specifically, multiplying $\Delta _{\uparrow \downarrow }$ by $(1-\zeta )$, we find that the gap is given by $\Delta \protect \sqrt {2} \protect \sqrt {1 + \zeta (\zeta /2 - 1)}$, which is $\protect \neq 0$ for all $\zeta \in (0,1)$.}\BibitemShut {Stop}%
\end{thebibliography}%

\end{document}